%% file: main.tex
\documentclass[fleqn]{2023SCGE}
\usepackage{amsmath}
\usepackage{aas_macros}
\usepackage{graphicx}
\usepackage{seqsplit}
\usepackage{natbib}

\newcommand{\gray}{$\gamma$-ray\ }

\newcommand{\grays}{$\gamma$-rays\ }

\newcommand{\signif}[1]{${\sim}#1\nobreak\,\sigma$}

\begin{document}

\ensubject{subject}

\ArticleType{Article}
\SpecialTopic{SPECIAL TOPIC: }
\Year{2023}
\Month{January}
\Vol{66}
\No{1}
\DOI{??}
\ArtNo{000000}
\ReceiveDate{ }
\AcceptDate{ }

\title{Observation of  the \gray Emission from  W43  with LHAASO}{Observation of  the \gray Emission from  W43  with LHAASO}

\author{LHAASO Collaboration\footnote{Corresponding author: G.W. Wang (wangguangwei@mail.ustc.edu.cn), C.D. Gao (gcd@mail.sdu.edu.cn), Y.H. Yu (yuyh@ustc.edu.cn), R.Z. Yang (yangrz@ustc.edu.cn)} \\(The LHAASO Collaboration authors and affiliations are listed after the references.)}{}%

\AuthorMark{Zhen Cao}
\AuthorCitation{Zhen Cao, et al}






\abstract{
In this paper, we report the detection of the very-high-energy (VHE, $ 100{\rm\ GeV} < E < 100{\rm\ TeV} $) and ultra-high-energy (UHE, $E > 100\rm\ TeV$) \gray emissions from the direction of the young star-forming region W43, observed by the Large High Altitude Air Shower Observation (LHAASO). The extended \gray source was detected with a significance of \signif{16} by KM2A and \signif{17}  by WCDA, respectively. The angular extension of this \gray source is about 0.5 degrees, corresponding to a physical size of about 50 pc. We discuss the origin of the \gray emission and possible cosmic ray acceleration in the W43 region using multi-wavelength data. Our findings suggest that W43 is likely another young star cluster capable of accelerating cosmic rays (CRs) to at least several hundred TeV.
}

\keywords{\gray source, Cosmic rays, Star formation region}

\PACS{07.85.-m, 96.50.S-, 97.10.Bt}

\maketitle


\begin{multicols}{2}

\section{Introduction}\label{sec:intro}

W43 is a giant HII region situated in the inner Galaxy \citep{smith78}. It is regarded as a Galactic mini starburst due to its extremely high star formation rate, contributing approximately $5\%$--$10\%$ of the total star formation rate in the Milky Way \citep{motte03,luong11}. The center of W43, dubbed W43-main,  is an extensive HII region energized by a Wolf-Rayet and OB star cluster. \gray emissions have also been detected in this region in both the GeV and TeV bands. The H.E.S.S. telescope detected an extended TeV source, HESS J1848-018, with a radius of $0.25^\circ$ \citep{hgps}. \cite{fermi_pwn} also found a faint GeV point-like source, consistent with the extension $~0.3^\circ$ reported by \citet{fermi_old}, and attributes it to a potential pulsar wind nebula, although the powering pulsar has not yet been found. Recently, \cite{yang20} have found an extended \gray emission with a radius of about $0.6^{\circ}$ and a \gray photon index of about $2.3$.  Assuming a distance of $5.5$ kpc \citep{w43_distance}, the large extension, with a physical size of about $50$ pc,  and the hard \gray spectrum \citep{yang20} are very similar to other \gray bright young massive star clusters in our Galaxy \citep{aharonian19}. In this scenario, the \gray emission around W43 is produced by the interaction of Cosmic Rays (CRs) injected by W43 with ambient gas, leading to the belief that W43 harbors cosmic ray accelerators.

The first \gray source catalog recently released by the LHAASO collaboration \citep{lhaaso_catalog}  includes the ultra-high-energy (UHE, $E > 100\rm\ TeV$) source 1LHAASO J1848-0153u, which spatially coincides with the GeV emissions mentioned above. In this catalog, `u' indicates that the significance is above $4\,\sigma$ at $E > 100\ {\rm TeV}$. Thus, it is naturally postulated that W43 could even be a PeV CR accelerator.
In this study, we conducted a detailed data analysis using LHAASO KM2A and WCDA data on the W43 region. Due to the increased exposure and an improved understanding of the instruments’ response, we have achieved a much more significant detection, allowing for detailed morphology and spectral study in this region.

This paper is organized as follows: In Sec.2 we describe the analysis of LHAASO data taking from both WCDA and KM2A arrays; in Sec.3 we discuss the possible origin of the \gray emission and implications of the obtained results; and in Sec.4, we provide a summary of our findings.

\section{LHAASO DATA ANALYSIS AND RESULTS} \label{sec:LDAAR}

\subsection{Data and analysis technique} \label{subsec:DAAT}

\begin{figure*}[ht]
    \centering
    \includegraphics[width=\linewidth]{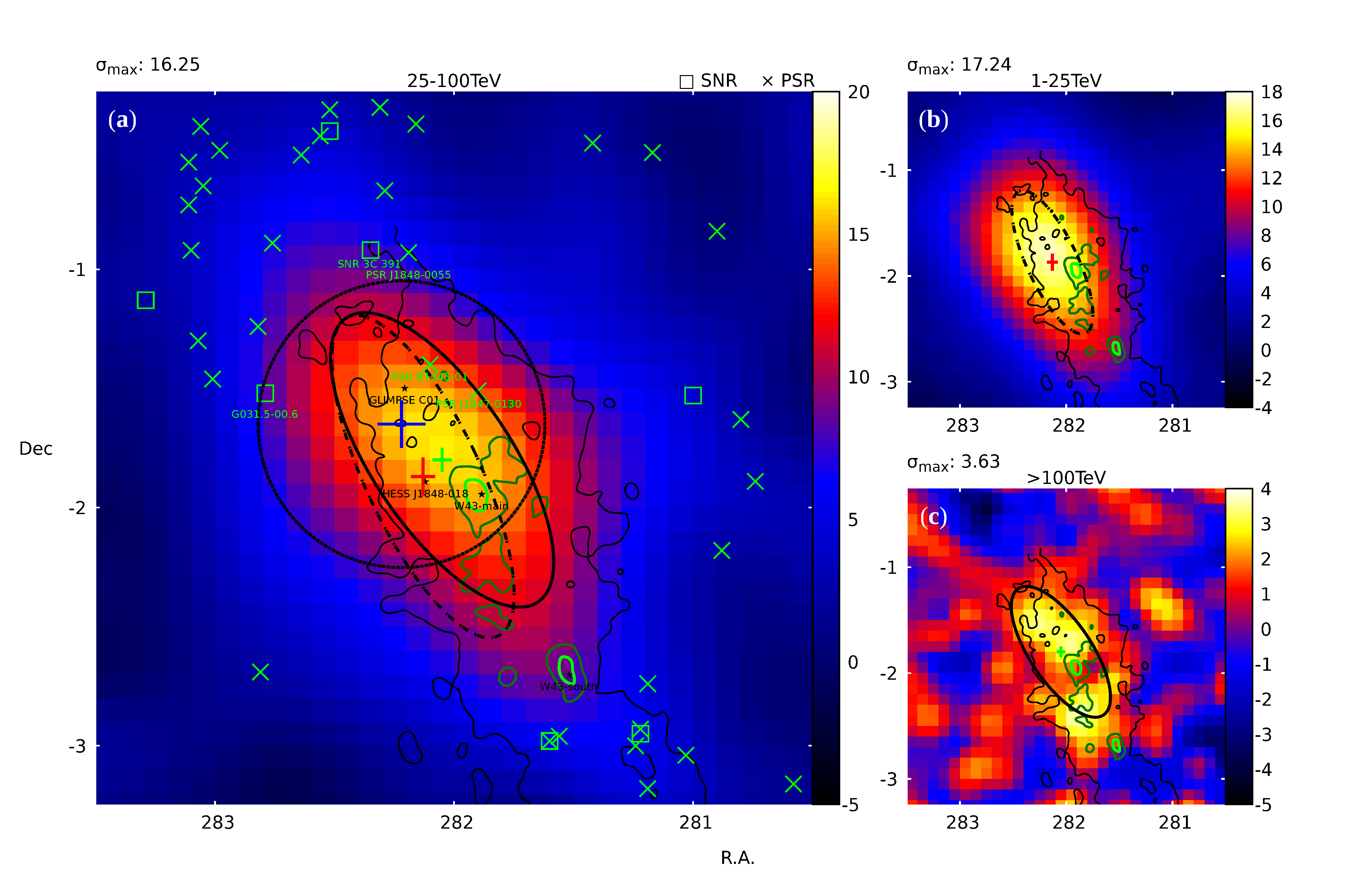}
        \caption{
             These figures display significance maps of W43 derived from the best-fit elliptical model. All sources, except LHAASO J1848-0153u, are subtracted with their best-fit parameters.  Figure (a) shows the significance map in the energy range from 25 TeV to 100 TeV. The green and red `+' symbols indicate the best-fit positions of KM2A and WCDA, respectively, along with their corresponding $1\,\sigma$ statistical error bars. The solid ellipse represents the best-fit morphology of LHAASO J1848-0153u as determined by KM2A, while the dot-dash ellipse corresponds to that of WCDA. The parameters used to draw these elements are taken from the best-fit elliptical models with spectral parameters of GDE left free. Similarly, the blue `+' symbol and the dashed circle show results from the Fermi data  \citep{yang20}. The contour lines in black, dark green and bright green indicate gas column densities of $ 3$, $6$ and $ 10\times 10^{22} {\rm\ cm^{-2}} $ \citep{yang20}, respectively. Supernova Remnants (SNRs, marked by square symbols) from \citet{greensnr1} and pulsars (PSRs, marked by cross symbols) from ATNF pulsar catalogue \citep{atnf1,atnf2} are also shown here. The positions of the two molecular clouds W43-main and W43-south, the TeV source HESS J1848-018 and the globular cluster GLIMPSE C01 are also plotted. Figure (b) shows the significance map in the energy range from 1 TeV to 25 TeV\@. Figure (c) shows the significance map in the energy range above 100 TeV\@.
        }
    \label{fig:W43map}
\end{figure*}

The Large High Altitude Air Shower Observatory (LHAASO) is located at Haizi Mountain, Daocheng, Sichuan province, China. It is composed of three sub-arrays, consisting of the $1\,{\rm km}^2$ array (KM2A), the Water Cherenkov Detector Array (WCDA), and the Wide Field-of-view air Cherenkov/fluorescence Telescope Array (WFCTA).

The \gray data  ($1\ {\rm TeV}<E< 25\ {\rm TeV}$ ) were collected by WCDA, which covers a physical area of $78,000\ \rm{m}^2$ \citep{Aharonian_2020}.  The data used in this study were obtained from the full-array operation of the WCDA from March 5, 2021, to May 31, 2023, with a total exposure time of 735 days. To select events, the $\gamma$-Proton identification parameter called "pincness"  \citep{osti_22876076} was set to less than 1.1, and the zenith angle of the shower was required to be less than $50^\circ$. The collected data were divided into five energy segments based on the number of triggered detectors (nhit), categorized into the following ranges: 100--200, 200--300, 300--500, 500--800, and 800--2000. The mean energy corresponding to each nhit bin is approximately 2, 4, 6, 11 and 23 TeV, respectively, with a power-law spectrum index of -2.6. For each energy segment, the sky map in celestial coordinates (right ascension and declination) was divided into a grid of $0.1 ^ \circ \times  0.1 ^ \circ $, where each pixel was filled with the number of the detected events according to their reconstructed arrival direction.  The cosmic ray background was estimated using the "Time-Swap" method \citep{Fleysher_2004}, with the swapping repeated 200 times for each real event.

The \gray data ($25\ {\rm TeV}<E< 4\ {\rm PeV}$)  were collected from different phases of the KM2A array: half the array from December 2019 to December 2020, three-quarters of the array from December 2020 to July 2021, and the full KM2A array from July 2021 to July 2023. After the data quality check \citep{quality_check}, the total effective observation time was 1216.23 days. The same event reconstruction and selection were used as described in the performance paper \citep{crabcpc}. The KM2A data sets were divided into five energy bins per decade according to the reconstructed energy. For the data set in each energy bin, the sky map was binned in the same way as WCDA. The “direct integration method”  \citep{Fleysher_2004} was used to estimate the cosmic-ray background.  In each bin, the number of observed events and the estimated background event number are simply the sums of those from different phases.
 
A 3D likelihood fitting process was applied to determine the spectrum and morphology of sources based on the binned likelihood in this analysis. The Test Statistic (TS) is used to compare the goodness of each hypothesis. It is defined as 
\begin{equation}
 {\rm TS} = \frac{\mathcal{L}_{s+b}}{\mathcal{L}_{b}} = \frac{\displaystyle \max \prod_{i}^{}{\rm Poisson}(N_i^{\rm obs}, \sum_{j}^{}{N_i^{\rm src_j}} + N_i^{\rm bkg})}{\displaystyle \max \prod_{i=1}^{n}{\rm Poisson}(N_i^{\rm obs}, N_i^{\rm bkg})}
\tag{1}.
\end{equation}
Here, ${\mathcal{L}_{s+b}}$ is the maximum likelihood value or the hypothesis that includes both the sources and the background, while ${\mathcal{L}_{b}}$ is for the background-only hypothesis. `i' is the index of each bin with specific energy and position, and `j' is the index for sources. $N_i^{\rm obs}$ is the observed event number in each bin. $N_i^{\rm bkg}$ is the estimated cosmic-ray background in each bin. $N_i^{\rm src_j}$ is the expected event number from each source in each bin taking into account the detector response. The Poisson term represents the statistical probability of the observed number of events. The TS value, according to Wilks' Theorem \citep{wilks}, follows a chi-square distribution. The number of degrees of freedom for this distribution corresponds to the difference in the number of degrees of freedom between the null hypothesis (background only) and the alternative hypothesis (source plus background).

To check for the presence of additional sources and to obtain the morphology or significance distribution of the observed sources, we generated TS maps using the maximum likelihood method. We conducted a hypothesis test at the center of each position bin. The null hypothesis assumes only background is present in the Region of Interest (ROI), while the alternative hypothesis suggests an additional point source in the test position. Specifically, a 2-dimensional Gaussian template is assumed for each position bin, with a spectral index of -2.6 for WCDA and -3 for KM2A, consistent with the first LHAASO catalog \citep{lhaaso_catalog}. If the TS value is greater than 25, or equivalently, if the significance is greater than 5, it suggests that there might be a signal excess at this location. The morphology of the significance distribution can also reflect the morphology of the source.

The ROI for the analysis is defined as a disk with a radius of 6$^{\circ}$, centered at the position of W43 ($\mathrm{RA} = 281.885^\circ$, $\mathrm{DEC} = -1.942^\circ$). To ascertain the presence of additional sources within the ROI, we used an iterative algorithm that employs 3D likelihood fitting. In each iteration, a free circular Gaussian source with a power-law spectrum is added to the ROI, until the increment of TS value of the new model is lower than 25. Thus, the last model whose increment of TS is higher than 25 is the best-fit model in the iterative procedure. Then we can obtain the parameters of morphology and spectrum for each source from the best-fit results.

\begin{table*}[]
    \footnotesize
    \begin{threeparttable}
        \caption{Best-fit parameters of LHAASO J1848-0153u for elliptical models}
        \doublerulesep 0.1pt \tabcolsep 13pt 
            \begin{tabular}{cccccccc}   
                \toprule
                \hline
                    Array &                R.A. &             Dec  &     $\sigma_x$ &     $\sigma_y$ &  $\theta$  &            $J $ &     $\alpha$ \\
                
                               &         $ ^\circ $  &      $ ^\circ $  &   $ ^\circ $             &    $ ^\circ $            &  $ ^\circ $   &           \tnote{1)} &                    \\                              
                \hline
                
                KM2A  &  $282.05\pm0.04$  &  $-1.80\pm0.05$  &  $0.29\pm0.05$  &  $0.72\pm0.08$  & $-34\pm4$ &  $0.23\pm0.02$  &  $3.70\pm0.11$  \\ 
                WCDA  &  $282.13\pm0.05$  &  $-1.87\pm0.08$  &  $0.24\pm0.04$  &  $0.74\pm0.12$  & $-25\pm4$ &  $0.50\pm0.07$  &  $2.69\pm0.08$  \\
        
                \hline
            \end{tabular}
            \label{tbl:bestpara}              
            \begin{tablenotes}
                \item[1)]$10^{-15}\,{\rm\ TeV}^{-1}\,{\rm cm}^{-2}\,{\rm s}^{-1} $ for KM2A, $10^{-13}\,{\rm\ TeV}^{-1}\,{\rm cm}^{-2}\,{\rm s}^{-1} $ for WCDA.
            \end{tablenotes}
    \end{threeparttable}
\end{table*}

\subsection{Results} \label{subsec:Results}
After applying the aforementioned procedure, we derived the residual significance map. This map shows the residual significance after subtracting all sources except for LHAASO J1848-0153u. The subtraction process was performed in accordance with the best-fit model. This process unveiled excess emissions in the vicinity of W43, as depicted in Figure \ref{fig:W43map}. The significance of signal excess reaches up to \signif{17} in the energy range from $1$~TeV to $25$~TeV, \signif{16} from $25$~TeV to $100$~TeV, and $3.6\ \sigma $ above $100~\rm TeV$.

We used the iterative algorithm described previously to study the morphology and the spectrum of the excess emission. We found that a circular Gaussian distribution alone was insufficient to explain the observations. Therefore, we explored alternative models, including one with two circular Gaussian sources near W43. This model was taken into consideration if its TS value exceeded that of the single circular Gaussian model by at least 25 in the iterative procedure. In another model, we replaced the circular Gaussian source with an elliptical Gaussian source that spatially coincides with the position of W43 during the iterative procedure. In the elliptical Gaussian model, we hypothesize the morphology of  LHAASO J1848-0153u  follows an elliptical Gaussian distribution, represented by: 
\begin{equation}
\frac{d\,P}{d\,\Omega} = \frac{1}{2\pi \sigma_x\sigma_y} \exp{(-a\cdot{\theta_x^2}+2b\cdot{\theta_x}{\theta_y} - c\cdot{\theta_y^2})} \tag{2a}
\end{equation}, where
\begin{align*}
a &= \cos{\theta^2}/{2\sigma_x^2} + \sin{\theta^2}/{2\sigma_y^2} \\ 
b &= -\sin{2\theta}/{4\sigma_x^2} + \sin{2\theta}/{4\sigma_y^2} \tag{2b}\\ 
c &= \sin{\theta^2}/{2\sigma_x^2} + \cos{\theta^2}/{2\sigma_y^2}
\label{equ:elliptical}.
\end{align*}
Here, P is the event probability. $\Omega$ is the solid angle. $\theta$ is the rotation angle.  $\sigma_x$ and $\sigma_y$ are the widths of the corresponding directions. $\theta_x$ and $\theta_y$ are the offsets to the center of source $\rm(R.A.,\ Dec)$. The spectrum of LHAASO J1848-0153u was always assumed to be a power law $ f(E) = J\cdot{(E/E_0)}^\alpha$ for different morphology models. The reference energy was chosen to be $50 {\rm\ TeV}$ for KM2A, and $7 {\rm\ TeV}$ for WCDA. 

To account for the contribution from Galactic Diffuse Emission (GDE) in our ROI, we added a spatial template based on the gas distribution.  Following the approach of \citet{lhaaso_catalog}, we assumed the diffuse CRs are uniform in the ROI, and the GDE is proportional to the  gas column density map which is  derived  from the PLANCK dust opacity map   \citep{planck_dust1, planck_dust2}. And the spectral parameters of GDE were left free in the likelihood fitting.

We calculated the Akaike Information Criterion (${\rm AIC}$) \citep{aic} values to compare the goodness of fit among the elliptical Gaussian, the single circular Gaussian, and the two circular Gaussian models. The ${\rm AIC}$ is defined as ${\rm AIC} = -2\log(\mathcal{L}) + 2k$, where $\mathcal{L}$ is the maximum likelihood value of the model and $k$ is the number of free parameters in the model. For KM2A data, we found that two sources cannot be resolved, and the ${\rm AIC}$ value of the best-fit elliptical model is 24.0 lower than that of the single Gaussian model. While for WCDA data, we can resolve two sources and the ${\rm AIC}$ value of the elliptical model is larger by 2.7  compared with that of the double Gaussian model. Thus, the KM2A data favor the elliptical Gaussian model, while for WCDA the current data can hardly distinguish these two models. For simplicity we adopted the elliptical Gaussian model as our spatial template in this work. 
However, we acknowledge the possibility that future observations with better angular resolution may resolve the \gray emission in this region into more than one source. The best-fit parameters for the elliptical Gaussian template are listed in Table \ref{tbl:bestpara}. The derived SEDs are shown in Figure \ref{fig:sed}.

In addition to the statistical uncertainties, there are various systematic uncertainties in measuring the SEDs. The systematic uncertainties associated with the detectors  align with the findings of \citet{crabcpc}. Variability in the operational detector units over time, due to a debugging process involving  a small percentage of the array, marginally impacts the flux measurements. The arrangement of the array also influences the discrimination of $\gamma$-ray from the background noise. Additionally, the atmospheric models used in simulations introduce a degree of systematic uncertainty, thereby affecting the efficiency of detection. Collectively, these factors contribute to an approximate $7\%$ systematic uncertainty in flux for KM2A. Applying the same methodology, the overall systematic uncertainty can be as large as 8\% on the flux for WCDA measurement \citep{lhaaso_catalog}. 
 
The contribution of systematic uncertainties from the CR background estimate method should also be considered. \citet{KM2A_diffuse_prl} studied the impact of  different time windows and different large scale efficiency correction parameters for CR background estimation, deriving  about $5\%$ systematic uncertainties from the CR background estimation method for the measured GDE flux in the inner Galaxy region ($|b| < 5^\circ$, $15^\circ <  l < 125^\circ$). Since the ROI in this analysis is much smaller than that used in the GDE analysis \citep{KM2A_diffuse_prl}, the uncertainties from the large scale CR background should also be  smaller than $5\%$ in our work. As a conservative estimate, we add an additional $5\%$ uncertainties to account for systematic errors in CR background estimation in both WCDA and KM2A bands.

Furthermore, the GDE can also introduce  significant systematic errors. The GDE  flux measured in \citet{KM2A_diffuse_prl} is indeed significantly higher than expected assuming a unified  CR distribution in the Galaxy. This excess might be due to a higher CR 'sea' density in the inner Galaxy compared to the solar neighborhood. This increased density could result from a higher source density in that region. Alternatively, another explanation for the enhanced GDE measured by LHAASO could be substantial contributions from gamma-ray emissions from sources. Indeed, recently the LHAASO collaboration have found a very extended \gray bubble in Cygnus region, which is believed to be illuminated by CRs injected by the CR accelerators in Cygnus region \citep{LHAASO_cygnus_2} and confined near the accelerators. If such very extended \gray structure is common in the Galactic plane, the GDE measured can be significantly contributed by such sources rather than by the contribution from CR ‘sea’.  Thus to test the possible influence of our measured SEDs on W43, in addition to the GDE template we used above (dust opacity spatial template and free spectral parameters), we also manually adjusted the normalization of GDE template to be $50\%$ and $150\%$ of the best-fit value in both the WCDA and KM2A analysis. These tests resulted in a flux variation of $\sim 35\%$ for KM2A and $\sim 50\%$ for WCDA.  The spatial extension of the \gray emission is also changed by $\sim 30\%$. Such marginal influences of GDE could be due to two factors. One is the higher GDE measured by LHAASO than expected in the inner Galaxy \citep{KM2A_diffuse_prl}. The other is the fact that W43 is located in the densest region in the Galactic plane. The detailed discussion on the origin and uncertainties of GDE is beyond the scope of this paper and we left the uncertainties on the W43 spectra from the choices of GDE models as part of the systematic errors.   We summed all these systematic uncertainties in the grey error bars in Figure \ref{fig:sed}.

\section{DISCUSSIONS} \label{sec:discussion}

In lower energy band, extended \gray emission towards W43 has also been  detected by Fermi-LAT \citep{yang20}. In the GeV band, the emission is modeled as a disk with a radius of approximately $0.6^{\circ}$, centered at ($\mathrm{RA} = 282.22^{\circ}\pm 0.1^{\circ}$, $\mathrm{DEC} = -1.65^{\circ}\pm 0.1^{\circ}$). Thus, the GeV emission aligns well in both extension and position with the multi-TeV source observed by us. In the following discussion, we assume that the \gray emission from GeV band to more than $100~\rm TeV$ share the same origin.  The \grays can be produced in either hadronic or the leptonic processes. The gas distributions have been investigated in detail in \citet{yang20} by using the HI data from THOR survey \citep{Wang2019} taking into account the self-absorption and CO data from GRS  \citep{Jackson2006} and FUGIN survey \citep{Umemoto2017}. The total mass in the \gray emission region is estimated to be $3\times 10^6 ~\rm M_{\odot}$, and the average density is $140 ~\rm cm^{-3}$. 

\begin{figure*}[ht]
    \centering
    \includegraphics[scale=0.5]{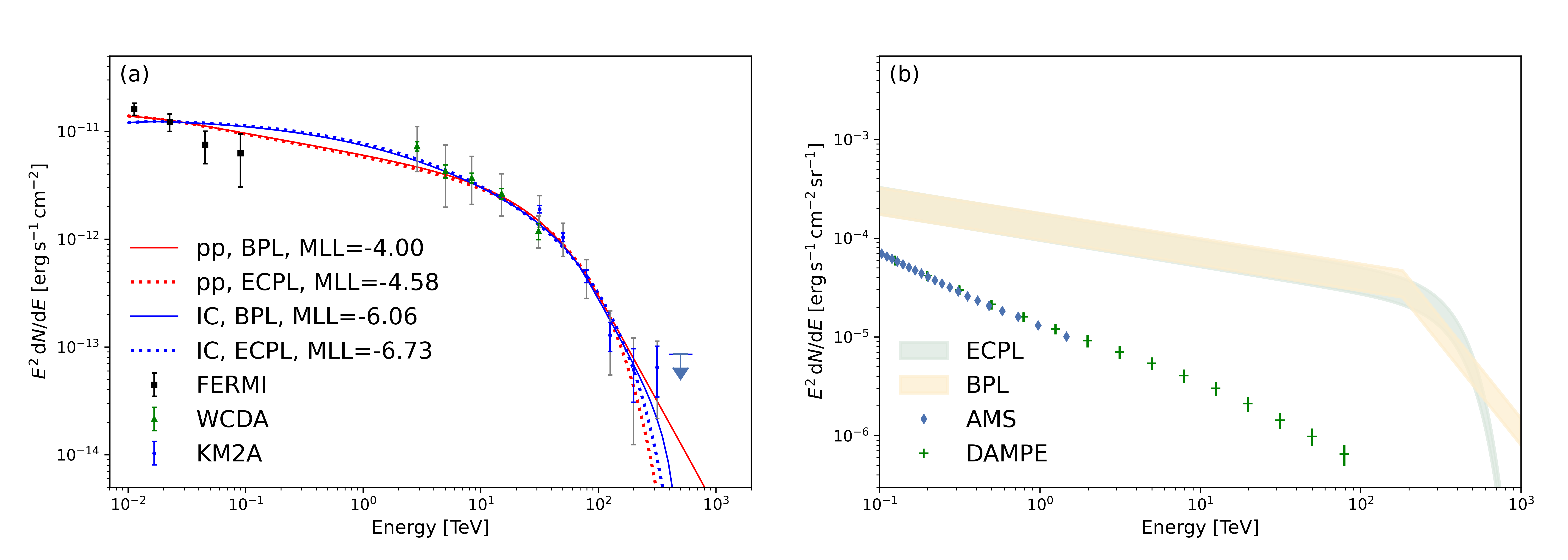}
    \caption{ Figure (a) shows the \gray spectra derived from the best-fit spectra of the parent particles, using the \textsf{NAIMA} package. Error bars with colors represent statistical errors, while the grey ones represent the systematic errors, primarily due to the challenges in understanding of the Galactic Diffuse Emission. The 95\% confidence upper limits are indicated with downward arrows. `MLL' represents the Maximum of Log(likelihood) in different models. In figure (a), the Fermi data points \citep{yang20} are also considered. In figure (b), we plotted bands by multiplying the best-fit proton spectra in both cases by 1.3 and 0.7, representing the uncertainties in gas density and pion decay cross-sections. They are also compared with the local proton spectrum measured by AMS \citep{ams_proton} and DAMPE \citep{dampe_proton}.}
    \label{fig:sed}
\end{figure*}

Figure \ref{fig:sed} shows the fit of the observed \gray spectra with the model  \citep{pp_production} in which CRs interact with ambient gas, using  {\textsf{NAIMA} \citep{naima}} package. In the fittings, we used the spectra derived in the previous section.  We discovered that both a broken power law (BPL) and a power law with an exponential cutoff (ECPL) are suitable models for parent protons. The formulae for BPL and ECPL are shown in Table \ref{tab:form}. The total CR energy budget ($>1\ {\rm TeV}$) is about $2.5\times 10^{48}~\rm erg$ assuming a gas density of about $140~\rm cm^{-3}$. The fitting results are shown in Table \ref{tab:sedfit1}. We note that in this case a quite sharp cutoff of break/cutoff in proton spectra is needed to fit the \gray flux. The derived CR density, as shown in Figure \ref{fig:sed} (b), is nearly 10 times higher than the CR density measured locally above 1 TeV.

Yet a leptonic origin cannot be excluded.  We also fit the SEDs in the leptonic scenario. In this scenario, the \gray emissions are assumed to be produced by the inverse Compton (IC) scattering of relativistic electrons with ambient photon fields \citep{naima_ic}. These photon fields include both the cosmic microwave background (CMB) and the interstellar radiation fields (ISRF) near W43 \citep{popescu17}. The total energy budget of electrons is derived to be $2.7 \times 10^{47}~ \rm erg$ above $1~\rm TeV$. This value is nearly one order of magnitude smaller than that in the hadronic case. To distinguish the radiation mechanism, the multi-wavelength observations, especially the X-ray measurements are extremely important. In this regard, the X-ray observations with the large field of view instruments, such as e-Rosita \citep{erosita} and Einstein Probe (EP) \citep{ep} may be particularly relevant.

W43 is a crowded region with several pulsars and SNRs coinciding with the \gray emission region.  These powerful objects  can be potential counterparts of the \gray sources. 
In addition, the globular cluster GLIMPSE C01 is located near the barycenter of the detected \gray emissions (as shown in Figure \ref{fig:W43map}). The dense region W43-south is also partially coinciding with the \gray emissions.

There are three pulsars in the \gray emission regions from the ATNF pulsar catalog \footnote{https://www.atnf.csiro.au} \citep{atnf1, atnf2}, which are  labeled in Figure \ref{fig:W43map} as PSR J1847-0130, PSR J1848-0055, and PSR B1845-01. Their spin-down luminosities are $1.7\times 10^{32} $, $2.6\times 10^{33} $, and $7.2\times 10^{32} \rm\,erg\,s^{-1}$, and their distances are 5.8, 7.4, and 4.4 kpc, respectively.  The relatively small spin-down luminosities make them rather unlikely to account for all the \gray emissions in this region. However, we cannot formally rule out the possibility that the \gray emissions are produced by the TeV halos associated with these pulsars \citep{lopez-coto22}. Specifically, PSR J1847-0130 has an age of about $8\times 10^4 $ years, which is similar to the age of PSR B0656+14, a known pulsar halo \citep{hawc_geminga}.

There are also several SNRs near this region, including SNR G31.5-0.6 which is located inside the \gray emission region. The SNR G31.5-0.6 is an SNR with an incomplete shell whose distance was estimated to be 12.9 kpc by using the $\Sigma-$D relation \citep{case98}.   In addition, another  mixed-morphology SNR 3C 391 is to the east of the \gray emission region. It is a mid-aged SNR with an age of about 4000 years and a distance of 7.2 kpc. The GeV emissions from this source are studied in detail in \citet{ergin14}.  The extended \grays in the W43 region might also result from interactions between ambient gas and CRs escaping from these SNRs. But the estimated distances imply that these SNRs are far from the dense gas in this region,  which disfavor the SNR scenario.

Indeed,  we cannot rule out the possibility that the emission in this region can be resolved into two sources with more exposure or instruments with better angular resolution. The WCDA data and the TS map above $100~\rm TeV$ show hints. In this scenario, the extended emission could be explained by the combination of two or several discrete components related to more compact emissions produced by pulsars or SNRs. But as shown in Figure \ref{fig:W43map} the two "hotspots" above $100~\rm TeV $ do not show strong correlation with pulsars/SNRs. Although in principle the \gray emissions can also be produced by CRs accelerated by their interaction with ambient gas.

The globular cluster  GLIMPSE C01 \citep{GLIMPSEC01} is also a possible astrophysical counterpart. H.E.S.S. collaboration have detected multi-TeV emissions from  globular cluster Terzan 5 \citep{hess_terzan5}, but the spatial extension is much smaller, which makes it quite unlikely that GLIMPSE C01 can contribute to all the \gray emissions in this region.

As discussed in \citet{yang20}, another plausible origin of the GeV emissions involves CRs accelerated in the W43 region interacting with ambient gas.  In this scenario, the detection of CRs up to several hundred TeV reveals the existence of PeV protons in this region. We also found a quite good correlation of the \gray map above 100 TeV with the gas distribution as shown in Figure \ref{fig:W43map}. This further supports the CRs interacting with ambient gas as the source of \gray emissions. We found an offset of about $0.2^\circ$ between the best-fit centroid of \gray emission and center of the massive cluster W43-main. In the current scenario that the CRs are injected by W43-main, such an offset is natural. Firstly, the acceleration/injection site of CRs is not necessarily located at the center of the cluster, as suggested by \cite{vieu22}. Secondly, the distribution of \gray emissions also depends on the distribution of gas.

\begin{table*}[]
    \centering
    \footnotesize
    \begin{threeparttable}
        \caption{Formulae for parent particle distribution used in the SED fitting process}\label{tab:sedfitfun}
        \doublerulesep 0.1pt \tabcolsep 13pt 
        \begin{tabular}{ccc}
                \toprule
                \hline
                Model Name  &  Formula\tnote{1)}                                                             &  Free parameters   \\ 
                \hline  
                ECPL  &  $N(E) = A (E/E_ 0)^{-\alpha} {\exp(-(E/E_\mathrm{cut})}^{\beta}) $  &  A, $\alpha$, $\beta$, $ E_\mathrm{cut}$ \\
                BPL   &  $N(E) =\begin{cases} A(E/E_0)^{-\alpha_1}                           &  \mbox{: }E<E_\mathrm{b} \\ A(E_\mathrm{b}/E_0)^{(\alpha_2-\alpha_1)}(E/E_0)^{-\alpha_2} & \mbox{: }E>E_\mathrm{b} \end{cases}$ &  A, $\alpha_1$, $\alpha_2$, $E_\mathrm{b}$\\
                \hline
                \bottomrule
        \end{tabular}
        \label{tab:form}
        \begin{tablenotes}
            \item[1)] {\bf $E_0 = 10 \rm\ TeV$}
        \end{tablenotes}
    \end{threeparttable}

   \hskip 10pt

    \footnotesize
    \begin{threeparttable}
        \caption{Best-fit results of the parent particles for LHAASO J1848-0153u \tnote{1)}}\label{tab:sedfitfun}
        \doublerulesep 0.1pt \tabcolsep 13pt 
        \begin{tabular}{ccccccc}
                \toprule
                \hline
                Distribution  &  $\alpha/\alpha_1$       &  $\beta/\alpha_2$        &  $E_{\rm cut}/E_{\rm b}$   &  $W_p(>1 ~\rm TeV)\tnote{2)}$                 &  ${\rm MLL}\tnote{3)}$  \\
                \hline              
ECPL(PP)  &  $2.27^{+0.05}_{-0.06}$  &  $3.10^{+1.80}_{-1.56}$  &  $406^{+142}_{-108}$   &  $2.51^{+0.42}_{-0.37} \times 10^{48}$ erg  &  -4.58  \\
BPL(PP)  &  $2.27^{+0.05}_{-0.06}$  &  $4.41^{+0.92}_{-0.74}$  &  $211^{+96}_{-82}$   &  $2.51^{+0.40}_{-0.39} \times 10^{48}$ erg  &  -4.00  \\
ECPL(IC)  &  $2.98^{+0.08}_{-0.09}$  &  $2.08^{+2.16}_{-1.21}$  &  $228^{+389}_{-87}$   &  $2.70^{+0.34}_{-0.32} \times 10^{47}$ erg  &  -6.73  \\
BPL(IC)  &  $2.99^{+0.07}_{-0.07}$  &  $4.25^{+1.04}_{-0.75}$  &  $97^{+82}_{-48}$   &  $2.73^{+0.29}_{-0.30} \times 10^{47}$ erg  &  -6.06  \\
                \hline 
                \bottomrule
        \end{tabular}
        \begin{tablenotes}
            \item[1)] The gas density used in {\textsf{NAIMA}} package is $ n=140\,{\rm cm}^{-3}$. The distance of the source is assumed to be 5.5 kpc \citep{w43_distance}.
            \item[2)] The integrated  energy of the parent particles exceeding 1 TeV.
            \item[3)] Maximum of Log(likelihood).
        \end{tablenotes}
    \end{threeparttable}
    \label{tab:sedfit1}
\end{table*}

The CR energy budget required is approximately $2.5\times 10^{48}~\rm erg$  above $1~\rm TeV$ in the vicinity of W43. Considering a similar power law index down to lower energy, the total energy is of approximately $10^{49}~\rm erg$ above $10~\rm GeV$.  The physical size is about $50~\rm pc$.  Both values are similar to those of  Cygnus cocoon \citep{yang20}.  The total wind power of W43 is estimated as $10^{37}\rm\,erg\,s^{-1}$. Assuming the age of several Myrs for the central young cluster, the total kinetic energy is of the order $10^{51}~\rm erg$, which indicates that the acceleration efficiency of $1\%$ would suffice to account for the detected \gray emission.

A large extension,  a hard spectrum in the GeV band, and a break at TeV have also been found in the Cygnus Cocoon region \citep{LHAASO_cygnus_2}. These similarities suggest that W43 can be an analog of Cygnus cocoon, and thus another example of extreme CR accelerators associated with young massive clusters. However, a sharp spectral break is observed in W43, which is quite different from the gradual softening of \gray spectrum in Cygnus. Since current spectral fitting cannot distinguish the BPL or ECPL model, it is possible the break/cutoff results simply from the acceleration limit of the particle accelerators. Thus, W43 likely harbors accelerators capable of accelerating protons up to $\sim 200~\rm TeV$, with LHAASO KM2A measuring the spectrum in the cutoff region. In this scenario, W43 does not contribute significantly to PeV CRs in our Galaxy, a situation quite different from that of the Cygnus region.   

Another possible explanation for the CR spectral break is the propagation effect. With a continuous CR injection, the CR density follows $f(E) \sim \frac{Q(E)}{D(E)} = E^{-s-\delta}$,  where $Q(E) \sim E^{-s}$ describes the energy dependence of the injection spectrum, and $D(E) \sim E^{\delta}$ shows that from the diffusive transport.  A change in the energy dependence of the diffusion coefficient  can  induce a break in the propagated CR spectrum. Such a break is not observed in the interstellar medium (ISM).
However, the environment in W43 could  differ significantly  from that of ISM due to its much higher gas density and CR density.  The MHD turbulent cascade in the ISM can be damped out effectively and CRs can only scatter off the turbulence which is self-generated by the streaming instability. Such processes have been studied in detail in starburst galaxies by \citet{krumholz20}. The environment in W43 is similar to these starburst galaxies. In this case, the effective diffusion coefficient is energy-independent at lower energy (below some critical energy $E_{\rm br}$) and increases rapidly at higher energy. Thus, the propagated CR spectrum, as well as the \gray spectrum, would reveal a break in $E_{\rm br}$. However, whether this break can be as sharp as observed may require further investigation.

Recently, LHAASO detected an extended structure in the Cygnus region, far beyond the Cygnus cocoon \citep{LHAASO_cygnus_2}.  This phenomenon is naturally explained by CRs being injected by a central source, which then diffuse outward to occupy a larger volume.  However, no such structures have been found in the vicinity of W43. This absence may simply be due to W43 being much farther from us compared to the Cygnus region. As a result, the surface brightness is too low to be detected by LHAASO.

\section{CONCLUSIONS} \label{sec:conclusions}

In conclusion, LHAASO detected extended \gray emissions in the VHE to UHE band.  Although other origins cannot be formally excluded due to the crowded nature of this region, the most plausible origin of the \gray emission is the CRs accelerated by massive stars in W43 interacting with the dense gas in the vicinity. Assuming such an explanation, W43 would be the second star-forming region that harbors  multi-hundred TeV CR accelerators in addition to the Cygnus region. In the same context as the Cygnus region, one of the most likely accelerators is the young massive clusters in the W43 region. Future observations in this region, including high-angular-resolution  \gray measurements performed by Imaging Air Cherenkov Telescope arrays (IACTs), such as H.E.S.S. \citep{hess}, ASTRI \citep{astri}, CTA \citep{cta} and LACT \citep{lact}, and multi-wavelength observations in X-rays, would be crucial to understand the origin of \gray emissions in this region and corresponding particle acceleration mechanisms. 

\section*{Acknowledgements}

We would like to thank all staff members who work at the LHAASO site above 4400 meter above
the sea level year round to maintain the detector and keep the water recycling system, electricity power supply and other components of the experiment operating smoothly. We are grateful to
Chengdu Management Committee of Tianfu New Area for the constant financial support for research
with LHAASO data. We appreciate the computing and data service support provided by the National High Energy Physics Data Center for the data analysis in this paper. This research work is
supported by the following grants: The National Natural Science Foundation of China No.12393854, No.12175121, No.12393851, No.12393852, No.12393853, No.12205314, No.12105301, No.12305120, No.12261160362, No.12105294, No.U1931201, No.12375107, No.12173039, the Department of Science and Technology of Sichuan
Province, China No.24NSFSC2319, Project for Young Scientists in Basic Research of Chinese Academy
of Sciences No.YSBR-061, 
and in Thailand by the National Science and Technology Development Agency (NSTDA) and the National Research Council of Thailand (NRCT) under the High-Potential Research Team Grant Program
(N42A650868).

\InterestConflict{The authors declare that they have no conflict of interest.}




\bibliographystyle{aasjournal}
\bibliography{main}

\end{multicols}

\clearpage
\input{CollaborationAuthor}

\end{document}

%% file: CollaborationAuthor.tex
Zhen Cao$^{1,2,3}$,
F. Aharonian$^{4,5}$,
Axikegu$^{6}$,
Y.X. Bai$^{1,3}$,
Y.W. Bao$^{7}$,
D. Bastieri$^{8}$,
X.J. Bi$^{1,2,3}$,
Y.J. Bi$^{1,3}$,
W. Bian$^{9}$,
A.V. Bukevich$^{10}$
% ...
(The LHAASO Collaboration)
$^{1}$ Key Laboratory of Particle Astrophysics \& Experimental Physics Division \& Computing Center, Institute of High Energy Physics, Chinese Academy of Sciences, 100049 Beijing, China\\
$^{2}$ University of Chinese Academy of Sciences, 100049 Beijing, China\\
$^{3}$ TIANFU Cosmic Ray Research Center, Chengdu, Sichuan,  China\\
$^{4}$ Dublin Institute for Advanced Studies, 31 Fitzwilliam Place, 2 Dublin, Ireland \\
$^{5}$ Max-Planck-Institut for Nuclear Physics, P.O. Box 103980, 69029  Heidelberg, Germany\\
$^{6}$ School of Physical Science and Technology \&  School of Information Science and Technology, Southwest Jiaotong University, 610031 Chengdu, Sichuan, China\\
$^{7}$ School of Astronomy and Space Science, Nanjing University, 210023 Nanjing, Jiangsu, China\\
$^{8}$ Center for Astrophysics, Guangzhou University, 510006 Guangzhou, Guangdong, China\\
$^{9}$ Tsung-Dao Lee Institute \& School of Physics and Astronomy, Shanghai Jiao Tong University, 200240 Shanghai, China\\
$^{10}$ Institute for Nuclear Research of Russian Academy of Sciences, 117312 Moscow, Russia\\
% ...


%% file: main.bbl
\begin{thebibliography}{}
\expandafter\ifx\csname natexlab\endcsname\relax\def\natexlab#1{#1}\fi
\providecommand{\url}[1]{\href{#1}{#1}}
\providecommand{\dodoi}[1]{doi:~\href{http://doi.org/#1}{\nolinkurl{#1}}}
\providecommand{\doeprint}[1]{\href{http://ascl.net/#1}{\nolinkurl{http://ascl.net/#1}}}
\providecommand{\doarXiv}[1]{\href{https://arxiv.org/abs/#1}{\nolinkurl{https://arxiv.org/abs/#1}}}

\bibitem[{{Abeysekara} {et~al.}(2017{\natexlab{a}}){Abeysekara}, {Albert}, {Alfaro}, {Alvarez}, {{\'A}lvarez}, {Arceo}, {Arteaga-Vel{\'a}zquez}, {Ayala Solares}, {Barber}, {Bautista-Elivar}, {Becerril}, {Belmont-Moreno}, {BenZvi}, {Berley}, {Braun}, {Brisbois}, {Caballero-Mora}, {Capistr{\'a}n}, {Carrami{\~n}ana}, {Casanova}, {Castillo}, {Cotti}, {Cotzomi}, {Couti{\~n}o de Le{\'o}n}, {de la Fuente}, {De Le{\'o}n}, {DeYoung}, {Dingus}, {DuVernois}, {D{\'\i}az-V{\'e}lez}, {Ellsworth}, {Fiorino}, {Fraija}, {Garc{\'\i}a-Gonz{\'a}lez}, {Gerhardt}, {Gonz{\'a}lez Mun{\"o}z}, {Gonz{\'a}lez}, {Goodman}, {Hampel-Arias}, {Harding}, {Hernandez}, {Hernandez-Almada}, {Hinton}, {Hui}, {H{\"u}ntemeyer}, {Iriarte}, {Jardin-Blicq}, {Joshi}, {Kaufmann}, {Kieda}, {Lara}, {Lauer}, {Lee}, {Lennarz}, {Le{\'o}n Vargas}, {Linnemann}, {Longinotti}, {Raya}, {Luna-Garc{\'\i}a}, {L{\'o}pez-Coto}, {Malone}, {Marinelli}, {Martinez}, {Martinez-Castellanos}, {Mart{\'\i}nez-Castro}, {Mart{\'\i}nez-Huerta}, {Matthews}, {Miranda-Romagnoli},
  {Moreno}, {Mostaf{\'a}}, {Nellen}, {Newbold}, {Nisa}, {Noriega-Papaqui}, {Pelayo}, {Pretz}, {P{\'e}rez-P{\'e}rez}, {Ren}, {Rho}, {Rivi{\`e}re}, {Rosa-Gonz{\'a}lez}, {Rosenberg}, {Ruiz-Velasco}, {Salazar}, {Salesa Greus}, {Sandoval}, {Schneider}, {Schoorlemmer}, {Sinnis}, {Smith}, {Springer}, {Surajbali}, {Taboada}, {Tibolla}, {Tollefson}, {Torres}, {Ukwatta}, {Villase{\~n}or}, {Weisgarber}, {Westerhoff}, {Wisher}, {Wood}, {Yapici}, {Yodh}, {Younk}, {Zepeda}, \& {Zhou}}]{osti_22876076}
{Abeysekara}, A.~U., {Albert}, A., {Alfaro}, R., {et~al.} 2017{\natexlab{a}}, \apj, 843, 39, \dodoi{10.3847/1538-4357/aa7555}

\bibitem[{{Abeysekara} {et~al.}(2017{\natexlab{b}}){Abeysekara}, {Albert}, {Alfaro}, {Alvarez}, {{\'A}lvarez}, {Arceo}, {Arteaga-Vel{\'a}zquez}, {Avila Rojas}, {Ayala Solares}, {Barber}, {Bautista-Elivar}, {Becerril}, {Belmont-Moreno}, {BenZvi}, {Berley}, {Bernal}, {Braun}, {Brisbois}, {Caballero-Mora}, {Capistr{\'a}n}, {Carrami{\~n}ana}, {Casanova}, {Castillo}, {Cotti}, {Cotzomi}, {Couti{\~n}o de Le{\'o}n}, {De Le{\'o}n}, {De la Fuente}, {Dingus}, {DuVernois}, {D{\'\i}az-V{\'e}lez}, {Ellsworth}, {Engel}, {Enr{\'\i}quez-Rivera}, {Fiorino}, {Fraija}, {Garc{\'\i}a-Gonz{\'a}lez}, {Garfias}, {Gerhardt}, {Gonz{\'a}lez Mu{\~n}oz}, {Gonz{\'a}lez}, {Goodman}, {Hampel-Arias}, {Harding}, {Hern{\'a}ndez}, {Hern{\'a}ndez-Almada}, {Hinton}, {Hona}, {Hui}, {H{\"u}ntemeyer}, {Iriarte}, {Jardin-Blicq}, {Joshi}, {Kaufmann}, {Kieda}, {Lara}, {Lauer}, {Lee}, {Lennarz}, {Vargas}, {Linnemann}, {Longinotti}, {Luis Raya}, {Luna-Garc{\'\i}a}, {L{\'o}pez-Coto}, {Malone}, {Marinelli}, {Martinez}, {Martinez-Castellanos},
  {Mart{\'\i}nez-Castro}, {Mart{\'\i}nez-Huerta}, {Matthews}, {Miranda-Romagnoli}, {Moreno}, {Mostaf{\'a}}, {Nellen}, {Newbold}, {Nisa}, {Noriega-Papaqui}, {Pelayo}, {Pretz}, {P{\'e}rez-P{\'e}rez}, {Ren}, {Rho}, {Rivi{\`e}re}, {Rosa-Gonz{\'a}lez}, {Rosenberg}, {Ruiz-Velasco}, {Salazar}, {Salesa Greus}, {Sandoval}, {Schneider}, {Schoorlemmer}, {Sinnis}, {Smith}, {Springer}, {Surajbali}, {Taboada}, {Tibolla}, {Tollefson}, {Torres}, {Ukwatta}, {Vianello}, {Weisgarber}, {Westerhoff}, {Wisher}, {Wood}, {Yapici}, {Yodh}, {Younk}, {Zepeda}, {Zhou}, {Guo}, {Hahn}, {Li}, \& {Zhang}}]{hawc_geminga}
---. 2017{\natexlab{b}}, Science, 358, 911, \dodoi{10.1126/science.aan4880}

\bibitem[{{Acero} {et~al.}(2013){Acero}, {Ackermann}, {Ajello}, {Allafort}, {Baldini}, {Ballet}, {Barbiellini}, {Bastieri}, {Bechtol}, {Bellazzini}, {Bland ford}, {Bloom}, {Bonamente}, {Bottacini}, {Brandt}, {Bregeon}, {Brigida}, {Bruel}, {Buehler}, {Buson}, {Caliandro}, {Cameron}, {Caraveo}, {Cecchi}, {Charles}, {Chaves}, {Chekhtman}, {Chiang}, {Chiaro}, {Ciprini}, {Claus}, {Cohen-Tanugi}, {Conrad}, {Cutini}, {Dalton}, {D'Ammando}, {de Palma}, {Dermer}, {Di Venere}, {Silva}, {Drell}, {Drlica-Wagner}, {Falletti}, {Favuzzi}, {Fegan}, {Ferrara}, {Focke}, {Franckowiak}, {Fukazawa}, {Funk}, {Fusco}, {Gargano}, {Gasparrini}, {Giglietto}, {Giordano}, {Giroletti}, {Glanzman}, {Godfrey}, {Gr{\'e}goire}, {Grenier}, {Grondin}, {Grove}, {Guiriec}, {Hadasch}, {Hanabata}, {Harding}, {Hayashida}, {Hayashi}, {Hays}, {Hewitt}, {Hill}, {Horan}, {Hou}, {Hughes}, {Inoue}, {Jackson}, {Jogler}, {J{\'o}hannesson}, {Johnson}, {Kamae}, {Kawano}, {Kerr}, {Kn{\"o}dlseder}, {Kuss}, {Lande}, {Larsson}, {Latronico}, {Lemoine-Goumard},
  {Longo}, {Loparco}, {Lovellette}, {Lubrano}, {Marelli}, {Massaro}, {Mayer}, {Mazziotta}, {McEnery}, {Mehault}, {Michelson}, {Mitthumsiri}, {Mizuno}, {Monte}, {Monzani}, {Morselli}, {Moskalenko}, {Murgia}, {Nakamori}, {Nemmen}, {Nuss}, {Ohsugi}, {Okumura}, {Orienti}, {Orlando}, {Ormes}, {Paneque}, {Panetta}, {Perkins}, {Pesce-Rollins}, {Piron}, {Pivato}, {Porter}, {Rain{\`o}}, {Rando}, {Razzano}, {Reimer}, {Reimer}, {Reposeur}, {Ritz}, {Roth}, {Rousseau}, {Saz Parkinson}, {Schulz}, {Sgr{\`o}}, {Siskind}, {Smith}, {Spandre}, {Spinelli}, {Suson}, {Takahashi}, {Takeuchi}, {Thayer}, {Thayer}, {Thompson}, {Tibaldo}, {Tibolla}, {Tinivella}, {Torres}, {Tosti}, {Troja}, {Uchiyama}, {Vandenbroucke}, {Vasileiou}, {Vianello}, {Vitale}, {Werner}, {Winer}, {Wood}, \& {Yang}}]{fermi_pwn}
{Acero}, F., {Ackermann}, M., {Ajello}, M., {et~al.} 2013, \apj, 773, 77, \dodoi{10.1088/0004-637X/773/1/77}

\bibitem[{{Actis} {et~al.}(2011){Actis}, {Agnetta}, {Aharonian}, {Akhperjanian}, {Aleksi{\'c}}, {Aliu}, {Allan}, {Allekotte}, {Antico}, {Antonelli}, \& et~al.}]{cta}
{Actis}, M., {Agnetta}, G., {Aharonian}, F., {et~al.} 2011, Experimental Astronomy, 32, 193, \dodoi{10.1007/s10686-011-9247-0}

\bibitem[{{Aguilar} {et~al.}(2015){Aguilar}, {Aisa}, {Alpat}, {Alvino}, {Ambrosi}, {Andeen}, {Arruda}, {Attig}, {Azzarello}, {Bachlechner}, {Barao}, {Barrau}, {Barrin}, {Bartoloni}, {Basara}, {Battarbee}, {Battiston}, {Bazo}, {Becker}, {Behlmann}, {Beischer}, {Berdugo}, {Bertucci}, {Bigongiari}, {Bindi}, {Bizzaglia}, {Bizzarri}, {Boella}, {de Boer}, {Bollweg}, {Bonnivard}, {Borgia}, {Borsini}, {Boschini}, {Bourquin}, {Burger}, {Cadoux}, {Cai}, {Capell}, {Caroff}, {Casaus}, {Cascioli}, {Castellini}, {Cernuda}, {Cerreta}, {Cervelli}, {Chae}, {Chang}, {Chen}, {Chen}, {Cheng}, {Chen}, {Cheng}, {Chou}, {Choumilov}, {Choutko}, {Chung}, {Clark}, {Clavero}, {Coignet}, {Consolandi}, {Contin}, {Corti}, {Gil}, {Coste}, {Creus}, {Crispoltoni}, {Cui}, {Dai}, {Delgado}, {Della Torre}, {Demirk{\"o}z}, {Derome}, {Di Falco}, {Di Masso}, {Dimiccoli}, {D{\'\i}az}, {von Doetinchem}, {Donnini}, {Du}, {Duranti}, {D'Urso}, {Eline}, {Eppling}, {Eronen}, {Fan}, {Farnesini}, {Feng}, {Fiandrini}, {Fiasson}, {Finch}, {Fisher},
  {Galaktionov}, {Gallucci}, {Garc{\'\i}a}, {Garc{\'\i}a-L{\'o}pez}, {Gargiulo}, {Gast}, {Gebauer}, {Gervasi}, {Ghelfi}, {Gillard}, {Giovacchini}, {Goglov}, {Gong}, {Goy}, {Grabski}, {Grandi}, {Graziani}, {Guandalini}, {Guerri}, {Guo}, {Haas}, {Habiby}, {Haino}, {Han}, {He}, {Heil}, {Hoffman}, {Hsieh}, {Huang}, {Huh}, {Incagli}, {Ionica}, {Jang}, {Jinchi}, {Kanishev}, {Kim}, {Kim}, {Kirn}, {Kossakowski}, {Kounina}, {Kounine}, {Koutsenko}, {Krafczyk}, {La Vacca}, {Laudi}, {Laurenti}, {Lazzizzera}, {Lebedev}, {Lee}, {Lee}, {Leluc}, {Levi}, {Li}, {Li}, {Li}, {Li}, {Li}, {Li}, {Li}, {Li}, {Li}, {Lim}, {Lin}, {Lipari}, {Lippert}, {Liu}, {Liu}, {Lolli}, {Lomtadze}, {Lu}, {Lu}, {Lu}, {Luebelsmeyer}, {Luo}, {Lv}, {Majka}, {Ma{\~n}{\'a}}, {Mar{\'\i}n}, {Martin}, {Mart{\'\i}nez}, {Masi}, {Maurin}, {Menchaca-Rocha}, {Meng}, {Mo}, {Morescalchi}, {Mott}, {M{\"u}ller}, {Ni}, {Nikonov}, {Nozzoli}, {Nunes}, {Obermeier}, {Oliva}, {Orcinha}, {Palmonari}, {Palomares}, {Paniccia}, {Papi}, {Pauluzzi}, {Pedreschi}, {Pensotti},
  {Pereira}, {Picot-Clemente}, {Pilo}, {Piluso}, {Pizzolotto}, {Plyaskin}, {Pohl}, {Poireau}, {Postaci}, {Putze}, {Quadrani}, {Qi}, {Qin}, {Qu}, {R{\"a}ih{\"a}}, {Rancoita}, {Rapin}, {Ricol}, {Rodr{\'\i}guez}, {Rosier-Lees}, {Rozhkov}, {Rozza}, {Sagdeev}, {Sandweiss}, {Saouter}, {Sbarra}, {Schael}, {Schmidt}, {von Dratzig}, {Schwering}, {Scolieri}, {Seo}, {Shan}, {Shan}, {Shi}, {Shi}, {Shi}, {Siedenburg}, {Son}, {Spada}, {Spinella}, {Sun}, {Sun}, {Tacconi}, {Tang}, {Tang}, {Tang}, {Tao}, {Tescaro}, {Ting}, {Ting}, {Tomassetti}, {Torsti}, {T{\"u}rko{\v{g}}lu}, {Urban}, {Vagelli}, {Valente}, {Vannini}, {Valtonen}, {Vaurynovich}, {Vecchi}, {Velasco}, {Vialle}, {Vitale}, {Vitillo}, {Wang}, {Wang}, {Wang}, {Wang}, {Wang}, {Wang}, {Weng}, {Whitman}, {Wienkenh{\"o}ver}, {Wu}, {Wu}, {Xia}, {Xie}, {Xie}, {Xiong}, {Xin}, {Xu}, {Xu}, {Yan}, {Yang}, {Yang}, {Ye}, {Yi}, {Yu}, {Yu}, {Zeissler}, {Zhang}, {Zhang}, {Zhang}, {Zhang}, {Zheng}, {Zhuang}, {Zhukov}, {Zichichi}, {Zimmermann}, {Zuccon}, {Zurbach}, \& {AMS
  Collaboration}}]{ams_proton}
{Aguilar}, M., {Aisa}, D., {Alpat}, B., {et~al.} 2015, \prl, 114, 171103, \dodoi{10.1103/PhysRevLett.114.171103}

\bibitem[{{Aharonian} {et~al.}(2019){Aharonian}, {Yang}, \& {de O{\~n}a Wilhelmi}}]{aharonian19}
{Aharonian}, F., {Yang}, R., \& {de O{\~n}a Wilhelmi}, E. 2019, Nature Astronomy, 3, 561, \dodoi{10.1038/s41550-019-0724-0}

\bibitem[{{Aharonian} {et~al.}(2006){Aharonian}, {Akhperjanian}, {Bazer-Bachi}, {Beilicke}, {Benbow}, {Berge}, {Bernl{\"o}hr}, {Boisson}, {Bolz}, {Borrel}, {Braun}, {Breitling}, {Brown}, {B{\"u}hler}, {B{\"u}sching}, {Carrigan}, {Chadwick}, {Chounet}, {Cornils}, {Costamante}, {Degrange}, {Dickinson}, {Djannati-Ata{\"\i}}, {O'C. Drury}, {Dubus}, {Egberts}, {Emmanoulopoulos}, {Espigat}, {Feinstein}, {Ferrero}, {Fiasson}, {Fontaine}, {Funk}, {Funk}, {Gallant}, {Giebels}, {Glicenstein}, {Goret}, {Hadjichristidis}, {Hauser}, {Hauser}, {Heinzelmann}, {Henri}, {Hermann}, {Hinton}, {Hofmann}, {Holleran}, {Horns}, {Jacholkowska}, {de Jager}, {Kh{\'e}lifi}, {Komin}, {Konopelko}, {Kosack}, {Latham}, {Le Gallou}, {Lemi{\`e}re}, {Lemoine-Goumard}, {Lohse}, {Martin}, {Martineau-Huynh}, {Marcowith}, {Masterson}, {McComb}, {de Naurois}, {Nedbal}, {Nolan}, {Noutsos}, {Orford}, {Osborne}, {Ouchrif}, {Panter}, {Pelletier}, {Pita}, {P{\"u}hlhofer}, {Punch}, {Raubenheimer}, {Raue}, {Rayner}, {Reimer}, {Reimer}, {Ripken}, {Rob},
  {Rolland}, {Rowell}, {Sahakian}, {Saug{\'e}}, {Schlenker}, {Schlickeiser}, {Schwanke}, {Sol}, {Spangler}, {Spanier}, {Steenkamp}, {Stegmann}, {Superina}, {Tavernet}, {Terrier}, {Th{\'e}oret}, {Tluczykont}, {van Eldik}, {Vasileiadis}, {Venter}, {Vincent}, {V{\"o}lk}, {Wagner}, \& {Ward}}]{hess}
{Aharonian}, F., {Akhperjanian}, A.~G., {Bazer-Bachi}, A.~R., {et~al.} 2006, \aap, 457, 899, \dodoi{10.1051/0004-6361:20065351}

\bibitem[{{Aharonian} {et~al.}(2020){Aharonian}, {Alekseenko}, {An}, {Axikegu}, {Bai}, {Bao}, {Bastieri}, {Bi}, {Cai}, {Cao}, {Cao}, {Chang}, {Chang}, {Chang}, {Chao}, {Chen}, {Chen}, {Chen}, {Chen}, {Chen}, {Chen}, {Chen}, {Chen}, {Chen}, {Chen}, {Chen}, {Chen}, {Cheng}, {Cheng}, {Cui}, {Cui}, {Cui}, {Dai}, {Dai}, {Dai}, {Danzengluobu}, {D'Ettorre Piazzoli}, {Fang}, {Fan}, {Fan}, {Feng}, {Feng}, {Feng}, {Feng}, {Gao}, {Gao}, {Gao}, {Ge}, {Geng}, {Gong}, {Gou}, {Gu}, {Guo}, {Guo}, {Han}, {He}, {He}, {Heller}, {He}, {He}, {Hou}, {Huang}, {Huang}, {Huang}, {Huang}, {Hu}, {Hu}, {Jia}, {Jiang}, {Ji}, {Jin}, {Ji}, {Levochkin}, {Liang}, {Liang}, {Li}, {Li}, {Li}, {Li}, {Li}, {Li}, {Li}, {Li}, {Li}, {Li}, {Li}, {Li}, {Li}, {Li}, {Li}, {Liu}, {Liu}, {Liu}, {Liu}, {Liu}, {Liu}, {Liu}, {Liu}, {Liu}, {Liu}, {Liu}, {Liu}, {Liu}, {Long}, {Lu}, {Lv}, {Ma}, {Ma}, {Mao}, {Masood}, {Ma}, {Mitthumsiri}, {Montaruli}, {Nan}, {Pattarakijwanich}, {Pei}, {Qiao}, {Qi}, {Ruffolo}, {Rulev}, {S{\'a}iz}, {Shao}, {Shchegolev}, {Sheng},
  {Shi}, {Stenkin}, {Stepanov}, {Sun}, {Tam}, {Tang}, {Tian}, {Volpe}, {Wang}, {Wang}, {Wang}, {Wang}, {Wang}, {Wang}, {Wang}, {Wang}, {Wang}, {Wang}, {Wang}, {Wang}, {Wang}, {Wang}, {Wang}, {Wang}, {Wang}, {Wei}, {Wei}, {Wen}, {Wu}, {Wu}, {Wu}, {Wu}, {Wu}, {Xiang}, {Xiao}, {Xin}, {Xing}, {Xu}, {Xue}, {Yan}, {Yang}, {Yang}, {Yang}, {Yang}, {Yang}, {Yang}, {Yao}, {Yao}, {Ye}, {Yin}, {Yin}, {You}, {You}, {Yuan}, {Yu}, {Jiang}, {Zeng}, {Zeng}, {Zeng}, {Zeng}, {Zha}, {Zhang}, {Zhang}, {Zhang}, {Zhang}, {Zhang}, {Zhang}, {Zhang}, {Zhang}, {Zhang}, {Zhang}, {Zhang}, {Zhang}, {Zhang}, {Zhang}, {Zhang}, {Zhao}, {Zhao}, {Zhao}, {Zhao}, {Zheng}, {Zheng}, {Zhou}, {Zhou}, {Zhou}, {Zhou}, {Zhu}, {Zhu}, {Zhu}, {Zhu}, {Zuo}, \& {LHAASO Collaboration}}]{Aharonian_2020}
{Aharonian}, F., {Alekseenko}, V., {An}, Q., {et~al.} 2020, Chinese Physics C, 44, 065001, \dodoi{10.1088/1674-1137/44/6/065001}

\bibitem[{{Aharonian} {et~al.}(2021){Aharonian}, {An}, {Axikegu}, {Bai}, {Bai}, {Bao}, {Bastieri}, {Bi}, {Bi}, {Cai}, {Cai}, {Cao}, {Cao}, {Chang}, {Chang}, {Chang}, {Chen}, {Chen}, {Chen}, {Chen}, {Chen}, {Chen}, {Chen}, {Chen}, {Chen}, {Chen}, {Chen}, {Chen}, {Chen}, {Cheng}, {Cheng}, {Cui}, {Cui}, {Cui}, {Dai}, {Dai}, {Dai}, {Danzengluobu}, {Della Volpe}, {Piazzoli}, {Dong}, {Fan}, {Fan}, {Fan}, {Fang}, {Fang}, {Feng}, {Feng}, {Feng}, {Feng}, {Gao}, {Gao}, {Gao}, {Gao}, {Ge}, {Geng}, {Gong}, {Gou}, {Gu}, {Guo}, {Guo}, {Guo}, {Guo}, {Han}, {He}, {He}, {He}, {He}, {He}, {He}, {Heller}, {Hor}, {Hou}, {Hou}, {Hu}, {Hu}, {Hu}, {Hu}, {Huang}, {Huang}, {Huang}, {Huang}, {Huang}, {Ji}, {Ji}, {Jia}, {Jiang}, {Jiang}, {Jin}, {Kuleshov}, {Levochkin}, {Li}, {Li}, {Li}, {Li}, {Li}, {Li}, {Li}, {Li}, {Li}, {Li}, {Li}, {Li}, {Li}, {Li}, {Li}, {Li}, {Li}, {Liang}, {Liang}, {Lin}, {Liu}, {Liu}, {Liu}, {Liu}, {Liu}, {Liu}, {Liu}, {Liu}, {Liu}, {Liu}, {Liu}, {Liu}, {Liu}, {Liu}, {Liu}, {Long}, {Lu}, {Lv}, {Ma}, {Ma}, {Ma},
  {Mao}, {Masood}, {Mitthumsiri}, {Montaruli}, {Nan}, {Pang}, {Pattarakijwanich}, {Pei}, {Qi}, {Ruffolo}, {Rulev}, {S{\'a}iz}, {Shao}, {Shchegolev}, {Sheng}, {Shi}, {Song}, {Stenkin}, {Stepanov}, {Sun}, {Sun}, {Sun}, {Tam}, {Tang}, {Tian}, {Wang}, {Wang}, {Wang}, {Wang}, {Wang}, {Wang}, {Wang}, {Wang}, {Wang}, {Wang}, {Wang}, {Wang}, {Wang}, {Wang}, {Wang}, {Wang}, {Wang}, {Wang}, {Wang}, {Wang}, {Wang}, {Wei}, {Wei}, {Wei}, {Wen}, {Wu}, {Wu}, {Wu}, {Wu}, {Wu}, {Xi}, {Xia}, {Xia}, {Xiang}, {Xiao}, {Xiao}, {Xin}, {Xin}, {Xing}, {Xu}, {Xu}, {Xue}, {Yan}, {Yang}, {Yang}, {Yang}, {Yang}, {Yang}, {Yang}, {Yang}, {Yao}, {Yao}, {Ye}, {Yin}, {Yin}, {You}, {You}, {Yu}, {Yuan}, {Zeng}, {Zeng}, {Zeng}, {Zeng}, {Zha}, {Zhai}, {Zhang}, {Zhang}, {Zhang}, {Zhang}, {Zhang}, {Zhang}, {Zhang}, {Zhang}, {Zhang}, {Zhang}, {Zhang}, {Zhang}, {Zhang}, {Zhang}, {Zhang}, {Zhang}, {Zhang}, {Zhang}, {Zhang}, {Zhao}, {Zhao}, {Zhao}, {Zhao}, {Zhao}, {Zheng}, {Zheng}, {Zhou}, {Zhou}, {Zhou}, {Zhou}, {Zhou}, {Zhou}, {Zhu}, {Zhu}, {Zhu},
  {Zhu}, {Zuo}, \& {(Lhaaso Collaboration)}}]{crabcpc}
{Aharonian}, F., {An}, Q., {Axikegu}, {et~al.} 2021, Chinese Physics C, 45, 025002, \dodoi{10.1088/1674-1137/abd01b}

\bibitem[{Akaike(1973)}]{aic}
Akaike, H. 1973, in Proceedings of the 2nd International Symposium on Information Theory, Budapest, 267--281

\bibitem[{{An} {et~al.}(2019){An}, {Asfandiyarov}, {Azzarello}, {Bernardini}, {Bi}, {Cai}, {Chang}, {Chen}, {Chen}, {Chen}, {Chen}, {Cui}, {Cui}, {Dai}, {D'Amone}, {De Benedittis}, {De Mitri}, {Di Santo}, {Ding}, {Dong}, {Dong}, {Dong}, {Donvito}, {Droz}, {Duan}, {Duan}, {D'Urso}, {Fan}, {Fan}, {Fang}, {Feng}, {Feng}, {Fusco}, {Gallo}, {Gan}, {Gao}, {Gargano}, {Gong}, {Gong}, {Guo}, {Guo}, {Guo}, {Han}, {Hu}, {Huang}, {Huang}, {Huang}, {Ionica}, {Jiang}, {Jin}, {Kong}, {Lei}, {Li}, {Li}, {Li}, {Li}, {Li}, {Liang}, {Liang}, {Liao}, {Liu}, {Liu}, {Liu}, {Liu}, {Liu}, {Liu}, {Loparco}, {Luo}, {Ma}, {Ma}, {Ma}, {Ma}, {Ma}, {Marsella}, {Mazziotta}, {Mo}, {Niu}, {Pan}, {Peng}, {Peng}, {Qiao}, {Rao}, {Salinas}, {Shang}, {Shen}, {Shen}, {Shen}, {Song}, {Su}, {Su}, {Sun}, {Surdo}, {Teng}, {Tykhonov}, {Vitillo}, {Wang}, {Wang}, {Wang}, {Wang}, {Wang}, {Wang}, {Wang}, {Wang}, {Wang}, {Wang}, {Wang}, {Wang}, {Wang}, {Wei}, {Wei}, {Wei}, {Wen}, {Wu}, {Wu}, {Wu}, {Wu}, {Wu}, {Xi}, {Xia}, {Xu}, {Xu}, {Xu}, {Xu}, {Xue},
  {Yang}, {Yang}, {Yang}, {Yang}, {Yao}, {Yu}, {Yuan}, {Yue}, {Zang}, {Zhang}, {Zhang}, {Zhang}, {Zhang}, {Zhang}, {Zhang}, {Zhang}, {Zhang}, {Zhang}, {Zhang}, {Zhang}, {Zhang}, {Zhang}, {Zhao}, {Zhao}, {Zhao}, {Zhou}, {Zhou}, {Zhu}, {Zhu}, \& {Zimmer}}]{dampe_proton}
{An}, Q., {Asfandiyarov}, R., {Azzarello}, P., {et~al.} 2019, Science Advances, 5, eaax3793, \dodoi{10.1126/sciadv.aax3793}

\bibitem[{{Cao} {et~al.}(2023){Cao}, {Aharonian}, {An}, {Axikegu}, {Bao}, {Bastieri}, {Bi}, {Bi}, {Cai}, {Cao}, {Cao}, {Cao}, {Chang}, {Chang}, {Chen}, {Chen}, {Chen}, {Chen}, {Chen}, {Chen}, {Chen}, {Chen}, {Chen}, {Chen}, {Chen}, {Chen}, {Cheng}, {Cheng}, {Cui}, {Cui}, {Cui}, {Cui}, {Dai}, {Dai}, {Dai}, {Danzengluobu}, {Dong}, {Duan}, {Fan}, {Fan}, {Fang}, {Fang}, {Feng}, {Feng}, {Feng}, {Feng}, {Feng}, {Gabici}, {Gao}, {Gao}, {Gao}, {Gao}, {Gao}, {Gao}, {Ge}, {Geng}, {Giacinti}, {Gong}, {Gou}, {Gu}, {Guo}, {Guo}, {Guo}, {Guo}, {Han}, {He}, {He}, {He}, {He}, {He}, {Heller}, {Hor}, {Hou}, {Hou}, {Hou}, {Hu}, {Hu}, {Hu}, {Huang}, {Huang}, {Huang}, {Huang}, {Huang}, {Huang}, {Huang}, {Ji}, {Jia}, {Jia}, {Jiang}, {Jiang}, {Jiang}, {Jin}, {Kang}, {Ke}, {Kuleshov}, {Kurinov}, {Li}, {Li}, {Li}, {Li}, {Li}, {Li}, {Li}, {Li}, {Li}, {Li}, {Li}, {Li}, {Li}, {Li}, {Li}, {Li}, {Li}, {Li}, {Li}, {Liang}, {Liang}, {Lin}, {Liu}, {Liu}, {Liu}, {Liu}, {Liu}, {Liu}, {Liu}, {Liu}, {Liu}, {Liu}, {Liu}, {Liu}, {Liu}, {Liu},
  {Lu}, {Luo}, {Lv}, {Ma}, {Ma}, {Ma}, {Mao}, {Min}, {Mitthumsiri}, {Mu}, {Nan}, {Neronov}, {Ou}, {Pang}, {Pattarakijwanich}, {Pei}, {Qi}, {Qi}, {Qiao}, {Qin}, {Ruffolo}, {S{\'a}iz}, {Semikoz}, {Shao}, {Shao}, {Shchegolev}, {Sheng}, {Shu}, {Song}, {Stenkin}, {Stepanov}, {Su}, {Sun}, {Sun}, {Sun}, {Tam}, {Tang}, {Tang}, {Tian}, {Wang}, {Wang}, {Wang}, {Wang}, {Wang}, {Wang}, {Wang}, {Wang}, {Wang}, {Wang}, {Wang}, {Wang}, {Wang}, {Wang}, {Wang}, {Wang}, {Wang}, {Wang}, {Wang}, {Wang}, {Wang}, {Wei}, {Wei}, {Wei}, {Wen}, {Wu}, {Wu}, {Wu}, {Wu}, {Wu}, {Xi}, {Xia}, {Xia}, {Xiang}, {Xiao}, {Xiao}, {Xin}, {Xin}, {Xing}, {Xiong}, {Xu}, {Xu}, {Xu}, {Xu}, {Xue}, {Yan}, {Yan}, {Yan}, {Yang}, {Yang}, {Yang}, {Yang}, {Yang}, {Yang}, {Yang}, {Yang}, {Yang}, {Yao}, {Yao}, {Ye}, {Yin}, {Yin}, {You}, {You}, {Yu}, {Yuan}, {Yue}, {Zeng}, {Zeng}, {Zeng}, {Zha}, {Zhang}, {Zhang}, {Zhang}, {Zhang}, {Zhang}, {Zhang}, {Zhang}, {Zhang}, {Zhang}, {Zhang}, {Zhang}, {Zhang}, {Zhang}, {Zhang}, {Zhang}, {Zhang}, {Zhang}, {Zhang}, {Zhao},
  {Zhao}, {Zhao}, {Zhao}, {Zhao}, {Zheng}, {Zhou}, {Zhou}, {Zhou}, {Zhou}, {Zhou}, {Zhou}, {Zhou}, {Zhu}, {Zhu}, {Zhu}, {Zhu}, {Zuo}, \& {Lhaaso Collaboration}}]{KM2A_diffuse_prl}
{Cao}, Z., {Aharonian}, F., {An}, Q., {et~al.} 2023, \prl, 131, 151001, \dodoi{10.1103/PhysRevLett.131.151001}

\bibitem[{Cao {et~al.}(2024{\natexlab{a}})Cao, Aharonian, An, Axikegu, Bai, Bao, Bastieri, Bi, Bi, Cai, Cao, Cao, Cao, Chang, Chang, Chen, Chen, Chen, Chen, Chen, Chen, Chen, Chen, Chen, Chen, Chen, Chen, Cheng, Cheng, Cui, Cui, Cui, Cui, Dai, Dai, Dai, Danzengluobu, della Volpe, Dong, Duan, Fan, Fan, Fang, Fang, Feng, Feng, Feng, Feng, Feng, Gabici, Gao, Gao, Gao, Gao, Gao, Gao, Ge, Geng, Giacinti, Gong, Gou, Gu, Guo, Guo, Guo, Guo, Han, He, He, He, He, He, Heller, Hor, Hou, Hou, Hou, Hu, Hu, Hu, Huang, Huang, Huang, Huang, Huang, Huang, Huang, Ji, Jia, Jia, Jiang, Jiang, Jiang, Jin, Kang, Ke, Kuleshov, Kurinov, Li, Li, Li, Li, Li, Li, Li, Li, Li, Li, Li, Li, Li, Li, Li, Li, Li, Li, Li, Liang, Liang, Lin, Liu, Liu, Liu, Liu, Liu, Liu, Liu, Liu, Liu, Liu, Liu, Liu, Liu, Liu, Lu, Luo, Lv, Ma, Ma, Ma, Mao, Min, Mitthumsiri, Mu, Nan, Neronov, Ou, Pang, Pattarakijwanich, Pei, Qi, Qi, Qiao, Qin, Ruffolo, Sáiz, Semikoz, Shao, Shao, Shchegolev, Sheng, Shu, Song, Stenkin, Stepanov, Su, Sun, Sun, Sun, Tam, Tang,
  Tang, Tian, Wang, Wang, Wang, Wang, Wang, Wang, Wang, Wang, Wang, Wang, Wang, Wang, Wang, Wang, Wang, Wang, Wang, Wang, Wang, Wang, Wang, Wei, Wei, Wei, Wen, Wu, Wu, Wu, Wu, Wu, Xi, Xia, Xia, Xiang, Xiao, Xiao, Xin, Xin, Xing, Xiong, Xu, Xu, Xu, Xu, Xue, Yan, Yan, Yan, Yang, Yang, Yang, Yang, Yang, Yang, Yang, Yang, Yang, Yao, Yao, Ye, Yin, Yin, You, You, Yu, Yuan, Yue, Zeng, Zeng, Zeng, Zha, Zhang, Zhang, Zhang, Zhang, Zhang, Zhang, Zhang, Zhang, Zhang, Zhang, Zhang, Zhang, Zhang, Zhang, Zhang, Zhang, Zhang, Zhang, Zhao, Zhao, Zhao, Zhao, Zhao, Zheng, Zhou, Zhou, Zhou, Zhou, Zhou, Zhou, Zhou, Zhu, Zhu, Zhu, Zhu, Zuo, \& Collaboration)}]{lhaaso_catalog}
Cao, Z., Aharonian, F., An, Q., {et~al.} 2024{\natexlab{a}}, The Astrophysical Journal Supplement Series, 271, 25, \dodoi{10.3847/1538-4365/acfd29}

\bibitem[{Cao {et~al.}(2024{\natexlab{b}})Cao, Aharonian, Axikegu, Bai, Bao, Bastieri, Bi, Bi, Bian, Bukevich, Cao, Cao, Cao, Chang, Chang, Chen, Chen, Chen, Chen, Chen, Chen, Chen, Chen, Chen, Chen, Chen, Chen, Chen, Chen, Cheng, Cheng, Cui, Cui, Cui, Cui, Dai, Dai, Dai, Danzengluobu, Dong, Duan, Fan, Fan, Fang, Fang, Fang, Feng, Feng, Feng, Feng, Feng, Feng, Feng, Gabici, Gao, Gao, Gao, Gao, Gao, Ge, Geng, Giacinti, Gong, Gou, Gu, Guo, Guo, Guo, Guo, Han, Hasan, He, He, He, He, Hor, Hou, Hou, Hou, Hu, Hu, Hu, Huang, Huang, Huang, Huang, Huang, Huang, Ji, Jia, Jia, Jiang, Jiang, Jiang, Jin, Kang, Karpikov, Kuleshov, Kurinov, Li, Li, Li, Li, Li, Li, Li, Li, Li, Li, Li, Li, Li, Li, Li, Li, Li, Li, Li, Liang, Liang, Lin, Liu, Liu, Liu, Liu, Liu, Liu, Liu, Liu, Liu, Liu, Liu, Liu, Liu, Liu, Luo, Luo, Lv, Ma, Ma, Ma, Mao, Min, Mitthumsiri, Mu, Nan, Neronov, Ou, Pattarakijwanich, Pei, Qi, Qi, Qiao, Qin, Raza, Ruffolo, Sáiz, Saeed, Semikoz, Shao, Shchegolev, Sheng, Shu, Song, Stenkin, Stepanov, Su, Sun, Sun, Sun,
  Sun, Takata, Tam, Tang, Tang, Tang, Tian, Wang, Wang, Wang, Wang, Wang, Wang, Wang, Wang, Wang, Wang, Wang, Wang, Wang, Wang, Wang, Wang, Wang, Wang, Wang, Wang, Wang, Wang, Wei, Wei, Wei, Wen, Wu, Wu, Wu, Wu, Wu, Wu, Xi, Xia, Xiang, Xiao, Xiao, Xin, Xing, Xiong, Xiong, Xu, Xu, Xu, Xu, Xue, Yan, Yan, Yan, Yang, Yang, Yang, Yang, Yang, Yang, Yang, Yang, Yao, Yao, Yin, Yin, You, You, Yu, Yuan, Yue, Zeng, Zeng, Zeng, Zha, Zhang, Zhang, Zhang, Zhang, Zhang, Zhang, Zhang, Zhang, Zhang, Zhang, Zhang, Zhang, Zhang, Zhang, Zhang, Zhang, Zhang, Zhang, Zhao, Zhao, Zhao, Zhao, Zhao, Zhao, Zheng, Zhong, Zhou, Zhou, Zhou, Zhou, Zhou, Zhou, Zhou, Zhou, Zhu, Zhu, Zhu, Zhu, Zhu, Zou, \& Zuo}]{quality_check}
Cao, Z., Aharonian, F., Axikegu, {et~al.} 2024{\natexlab{b}}, Data quality control system and long-term performance monitor of the LHAASO-KM2A.
\newblock \doarXiv{2405.11826}

\bibitem[{Cao {et~al.}(2024{\natexlab{c}})}]{LHAASO_cygnus_2}
Cao, Z., {et~al.} 2024{\natexlab{c}}, Sci. Bull., 69, 449, \dodoi{10.1016/j.scib.2023.12.040}

\bibitem[{{Case} \& {Bhattacharya}(1998)}]{case98}
{Case}, G.~L., \& {Bhattacharya}, D. 1998, \apj, 504, 761, \dodoi{10.1086/306089}

\bibitem[{{Ergin} {et~al.}(2014){Ergin}, {Sezer}, {Saha}, {Majumdar}, {Chatterjee}, {Bayirli}, \& {Ercan}}]{ergin14}
{Ergin}, T., {Sezer}, A., {Saha}, L., {et~al.} 2014, \apj, 790, 65, \dodoi{10.1088/0004-637X/790/1/65}

\bibitem[{{Fleysher} {et~al.}(2004){Fleysher}, {Fleysher}, {Nemethy}, {Mincer}, \& {Haines}}]{Fleysher_2004}
{Fleysher}, R., {Fleysher}, L., {Nemethy}, P., {Mincer}, A.~I., \& {Haines}, T.~J. 2004, \apj, 603, 355, \dodoi{10.1086/381384}

\bibitem[{{Green}(2019)}]{greensnr1}
{Green}, D.~A. 2019, Journal of Astrophysics and Astronomy, 40, 36, \dodoi{10.1007/s12036-019-9601-6}

\bibitem[{{H.~E.~S.~S. Collaboration} {et~al.}(2018){H.~E.~S.~S. Collaboration}, {Abdalla}, {Abramowski}, {Aharonian}, {Ait Benkhali}, {Ang{\"u}ner}, {Arakawa}, {Arrieta}, {Aubert}, {Backes}, {Balzer}, {Barnard}, {Becherini}, {Becker Tjus}, {Berge}, {Bernhard}, {Bernl{\"o}hr}, {Blackwell}, {B{\"o}ttcher}, {Boisson}, {Bolmont}, {Bonnefoy}, {Bordas}, {Bregeon}, {Brun}, {Brun}, {Bryan}, {B{\"u}chele}, {Bulik}, {Capasso}, {Carrigan}, {Caroff}, {Carosi}, {Casanova}, {Cerruti}, {Chakraborty}, {Chaves}, {Chen}, {Chevalier}, {Colafrancesco}, {Condon}, {Conrad}, {Davids}, {Decock}, {Deil}, {Devin}, {deWilt}, {Dirson}, {Djannati-Ata{\"\i}}, {Domainko}, {Donath}, {Drury}, {Dutson}, {Dyks}, {Edwards}, {Egberts}, {Eger}, {Emery}, {Ernenwein}, {Eschbach}, {Farnier}, {Fegan}, {Fernand es}, {Fiasson}, {Fontaine}, {F{\"o}rster}, {Funk}, {F{\"u}{\ss}ling}, {Gabici}, {Gallant}, {Garrigoux}, {Gast}, {Gat{\'e}}, {Giavitto}, {Giebels}, {Glawion}, {Glicenstein}, {Gottschall}, {Grondin}, {Hahn}, {Haupt}, {Hawkes}, {Heinzelmann},
  {Henri}, {Hermann}, {Hinton}, {Hofmann}, {Hoischen}, {Holch}, {Holler}, {Horns}, {Ivascenko}, {Iwasaki}, {Jacholkowska}, {Jamrozy}, {Jankowsky}, {Jankowsky}, {Jingo}, {Jouvin}, {Jung-Richardt}, {Kastendieck}, {Katarzy{\'n}ski}, {Katsuragawa}, {Katz}, {Kerszberg}, {Khangulyan}, {Kh{\'e}lifi}, {King}, {Klepser}, {Klochkov}, {Klu{\'z}niak}, {Komin}, {Kosack}, {Krakau}, {Kraus}, {Kr{\"u}ger}, {Laffon}, {Lamanna}, {Lau}, {Lees}, {Lefaucheur}, {Lemi{\`e}re}, {Lemoine-Goumard}, {Lenain}, {Leser}, {Lohse}, {Lorentz}, {Liu}, {L{\'o}pez-Coto}, {Lypova}, {Marandon}, {Malyshev}, {Marcowith}, {Mariaud}, {Marx}, {Maurin}, {Maxted}, {Mayer}, {Meintjes}, {Meyer}, {Mitchell}, {Moderski}, {Mohamed}, {Mohrmann}, {Mor{\r{a}}}, {Moulin}, {Murach}, {Nakashima}, {de Naurois}, {Ndiyavala}, {Niederwanger}, {Niemiec}, {Oakes}, {O'Brien}, {Odaka}, {Ohm}, {Ostrowski}, {Oya}, {Padovani}, {Panter}, {Parsons}, {Paz Arribas}, {Pekeur}, {Pelletier}, {Perennes}, {Petrucci}, {Peyaud}, {Piel}, {Pita}, {Poireau}, {Poon}, {Prokhorov},
  {Prokoph}, {P{\"u}hlhofer}, {Punch}, {Quirrenbach}, {Raab}, {Rauth}, {Reimer}, {Reimer}, {Renaud}, {de los Reyes}, {Rieger}, {Rinchiuso}, {Romoli}, {Rowell}, {Rudak}, {Rulten}, {Safi-Harb}, {Sahakian}, {Saito}, {Sanchez}, {Santangelo}, {Sasaki}, {Schand ri}, {Schlickeiser}, {Sch{\"u}ssler}, {Schulz}, {Schwanke}, {Schwemmer}, {Seglar-Arroyo}, {Settimo}, {Seyffert}, {Shafi}, {Shilon}, {Shiningayamwe}, {Simoni}, {Sol}, {Spanier}, {Spir-Jacob}, {Stawarz}, {Steenkamp}, {Stegmann}, {Steppa}, {Sushch}, {Takahashi}, {Tavernet}, {Tavernier}, {Taylor}, {Terrier}, {Tibaldo}, {Tiziani}, {Tluczykont}, {Trichard}, {Tsirou}, {Tsuji}, {Tuffs}, {Uchiyama}, {van der Walt}, {van Eldik}, {van Rensburg}, {van Soelen}, {Vasileiadis}, {Veh}, {Venter}, {Viana}, {Vincent}, {Vink}, {Voisin}, {V{\"o}lk}, {Vuillaume}, {Wadiasingh}, {Wagner}, {Wagner}, {Wagner}, {White}, {Wierzcholska}, {Willmann}, {W{\"o}rnlein}, {Wouters}, {Yang}, {Zaborov}, {Zacharias}, {Zanin}, {Zdziarski}, {Zech}, {Zefi}, {Ziegler}, {Zorn}, \&
  {{\.Z}ywucka}}]{hgps}
{H.~E.~S.~S. Collaboration}, {Abdalla}, H., {Abramowski}, A., {et~al.} 2018, \aap, 612, A1, \dodoi{10.1051/0004-6361/201732098}

\bibitem[{{H.E.S.S. Collaboration} {et~al.}(2011){H.E.S.S. Collaboration}, {Abramowski, A.}, {Acero, F.}, {Aharonian, F.}, {Akhperjanian, A. G.}, {Anton, G.}, {Balzer, A.}, {Barnacka, A.}, {Barres de Almeida, U.}, {Becherini, Y.}, {Becker, J.}, {Behera, B.}, {Bernlöhr, K.}, {Bochow, A.}, {Boisson, C.}, {Bolmont, J.}, {Bordas, P.}, {Brucker, J.}, {Brun, F.}, {Brun, P.}, {Bulik, T.}, {Büsching, I.}, {Carrigan, S.}, {Casanova, S.}, {Cerruti, M.}, {Chadwick, P. M.}, {Charbonnier, A.}, {Chaves, R. C. G.}, {Cheesebrough, A.}, {Chounet, L.-M.}, {Clapson, A. C.}, {Coignet, G.}, {Cologna, G.}, {Conrad, J.}, {Dalton, M.}, {Daniel, M. K.}, {Davids, I. D.}, {Degrange, B.}, {Deil, C.}, {Dickinson, H. J.}, {Djannati-Ataï, A.}, {Domainko, W.}, {Drury, L. O’C.}, {Dubois, F.}, {Dubus, G.}, {Dutson, K.}, {Dyks, J.}, {Dyrda, M.}, {Egberts, K.}, {Eger, P.}, {Espigat, P.}, {Fallon, L.}, {Farnier, C.}, {Fegan, S.}, {Feinstein, F.}, {Fernandes, M. V.}, {Fiasson, A.}, {Fontaine, G.}, {Förster, A.}, {Füßling, M.}, {Gallant,
  Y. A.}, {Gast, H.}, {Gérard, L.}, {Gerbig, D.}, {Giebels, B.}, {Glicenstein, J. F.}, {Glück, B.}, {Goret, P.}, {Göring, D.}, {Häffner, S.}, {Hague, J. D.}, {Hampf, D.}, {Hauser, M.}, {Heinz, S.}, {Heinzelmann, G.}, {Henri, G.}, {Hermann, G.}, {Hinton, J. A.}, {Hoffmann, A.}, {Hofmann, W.}, {Hofverberg, P.}, {Holler, M.}, {Horns, D.}, {Jacholkowska, A.}, {de Jager, O. C.}, {Jahn, C.}, {Jamrozy, M.}, {Jung, I.}, {Kastendieck, M. A.}, {Katarzyński, K.}, {Katz, U.}, {Kaufmann, S.}, {Keogh, D.}, {Khangulyan, D.}, {Khélifi, B.}, {Klochkov, D.}, {Kluźniak, W.}, {Kneiske, T.}, {Komin, Nu.}, {Kosack, K.}, {Kossakowski, R.}, {Laffon, H.}, {Lamanna, G.}, {Lennarz, D.}, {Lohse, T.}, {Lopatin, A.}, {Lu, C.-C.}, {Marandon, V.}, {Marcowith, A.}, {Masbou, J.}, {Maurin, D.}, {Maxted, N.}, {McComb, T. J. L.}, {Medina, M. C.}, {Méhault, J.}, {Nguyen, N.}, {Moderski, R.}, {Moulin, E.}, {Naumann, C. L.}, {Naumann-Godo, M.}, {de Naurois, M.}, {Nedbal, D.}, {Nekrassov, D.}, {Nicholas, B.}, {Niemiec, J.}, {Nolan, S. J.},
  {Ohm, S.}, {de Oña Wilhelmi, E.}, {Opitz, B.}, {Ostrowski, M.}, {Oya, I.}, {Panter, M.}, {Paz Arribas, M.}, {Pedaletti, G.}, {Pelletier, G.}, {Petrucci, P.-O.}, {Pita, S.}, {Pühlhofer, G.}, {Punch, M.}, {Quirrenbach, A.}, {Raue, M.}, {Rayner, S. M.}, {Reimer, A.}, {Reimer, O.}, {Renaud, M.}, {de los Reyes, R.}, {Rieger, F.}, {Ripken, J.}, {Rob, L.}, {Rosier-Lees, S.}, {Rowell, G.}, {Rudak, B.}, {Rulten, C. B.}, {Ruppel, J.}, {Ryde, F.}, {Sahakian, V.}, {Santangelo, A.}, {Schlickeiser, R.}, {Schöck, F. M.}, {Schulz, A.}, {Schwanke, U.}, {Schwarzburg, S.}, {Schwemmer, S.}, {Sikora, M.}, {Skilton, J. L.}, {Sol, H.}, {Spengler, G.}, {Stawarz, Ł.}, {Steenkamp, R.}, {Stegmann, C.}, {Stinzing, F.}, {Stycz, K.}, {Sushch, I.}, {Szostek, A.}, {Tavernet, J.-P.}, {Terrier, R.}, {Tluczykont, M.}, {Valerius, K.}, {van Eldik, C.}, {Vasileiadis, G.}, {Venter, C.}, {Vialle, J. P.}, {Viana, A.}, {Vincent, P.}, {Völk, H. J.}, {Volpe, F.}, {Vorobiov, S.}, {Vorster, M.}, {Wagner, S. J.}, {Ward, M.}, {White, R.},
  {Wierzcholska, A.}, {Zacharias, M.}, {Zajczyk, A.}, {Zdziarski, A. A.}, {Zech, A.}, \& {Zechlin, H.-S.}}]{hess_terzan5}
{H.E.S.S. Collaboration}, {Abramowski, A.}, {Acero, F.}, {et~al.} 2011, A\&A, 531, L18, \dodoi{10.1051/0004-6361/201117171}

\bibitem[{{Jackson} {et~al.}(2006){Jackson}, {Rathborne}, {Shah}, {Simon}, {Bania}, {Clemens}, {Chambers}, {Johnson}, {Dormody}, {Lavoie}, \& {Heyer}}]{Jackson2006}
{Jackson}, J.~M., {Rathborne}, J.~M., {Shah}, R.~Y., {et~al.} 2006, \apjs, 163, 145, \dodoi{10.1086/500091}

\bibitem[{{Kafexhiu} {et~al.}(2014){Kafexhiu}, {Aharonian}, {Taylor}, \& {Vila}}]{pp_production}
{Kafexhiu}, E., {Aharonian}, F., {Taylor}, A.~M., \& {Vila}, G.~S. 2014, \prd, 90, 123014, \dodoi{10.1103/PhysRevD.90.123014}

\bibitem[{{Khangulyan} {et~al.}(2014){Khangulyan}, {Aharonian}, \& {Kelner}}]{naima_ic}
{Khangulyan}, D., {Aharonian}, F.~A., \& {Kelner}, S.~R. 2014, \apj, 783, 100, \dodoi{10.1088/0004-637X/783/2/100}

\bibitem[{{Kobulnicky} {et~al.}(2005){Kobulnicky}, {Monson}, {Buckalew}, {Darnel}, {Uzpen}, {Meade}, {Babler}, {Indebetouw}, {Whitney}, {Watson}, {Churchwell}, {Wolfire}, {Wolff}, {Clemens}, {Shah}, {Bania}, {Benjamin}, {Cohen}, {Dickey}, {Jackson}, {Marston}, {Mathis}, {Mercer}, {Stauffer}, {Stolovy}, {Norris}, {Kutyrev}, {Canterna}, \& {Pierce}}]{GLIMPSEC01}
{Kobulnicky}, H.~A., {Monson}, A.~J., {Buckalew}, B.~A., {et~al.} 2005, \aj, 129, 239, \dodoi{10.1086/426337}

\bibitem[{{Krumholz} {et~al.}(2020){Krumholz}, {Crocker}, {Xu}, {Lazarian}, {Rosevear}, \& {Bedwell-Wilson}}]{krumholz20}
{Krumholz}, M.~R., {Crocker}, R.~M., {Xu}, S., {et~al.} 2020, \mnras, 493, 2817, \dodoi{10.1093/mnras/staa493}

\bibitem[{{Lemoine-Goumard} {et~al.}(2011){Lemoine-Goumard}, {Ferrara}, {Grondin}, {Martin}, \& {Renaud}}]{fermi_old}
{Lemoine-Goumard}, M., {Ferrara}, E., {Grondin}, M.~H., {Martin}, P., \& {Renaud}, M. 2011, \memsai, 82, 739, \dodoi{10.48550/arXiv.1109.4733}

\bibitem[{{L{\'o}pez-Coto} {et~al.}(2022){L{\'o}pez-Coto}, {de O{\~n}a Wilhelmi}, {Aharonian}, {Amato}, \& {Hinton}}]{lopez-coto22}
{L{\'o}pez-Coto}, R., {de O{\~n}a Wilhelmi}, E., {Aharonian}, F., {Amato}, E., \& {Hinton}, J. 2022, Nature Astronomy, 6, 199, \dodoi{10.1038/s41550-021-01580-0}

\bibitem[{{Manchester} {et~al.}(2005){Manchester}, {Hobbs}, {Teoh}, \& {Hobbs}}]{atnf1}
{Manchester}, R.~N., {Hobbs}, G.~B., {Teoh}, A., \& {Hobbs}, M. 2005, \aj, 129, 1993, \dodoi{10.1086/428488}

\bibitem[{Manchester {et~al.}(2023)Manchester, Hobbs, Teoh, \& Hobbs}]{atnf2}
Manchester, R.~N., Hobbs, G.~B., Teoh, A., \& Hobbs, M. 2023, {The Australia Telescope National Facility Pulsar Catalogue (1.70 version)}.
\newblock \url{https://www.atnf.csiro.au/research/pulsar/psrcat/}

\bibitem[{{Merloni} {et~al.}(2012){Merloni}, {Predehl}, {Becker}, {B{\"o}hringer}, {Boller}, {Brunner}, {Brusa}, {Dennerl}, {Freyberg}, {Friedrich}, {Georgakakis}, {Haberl}, {Hasinger}, {Meidinger}, {Mohr}, {Nandra}, {Rau}, {Reiprich}, {Robrade}, {Salvato}, {Santangelo}, {Sasaki}, {Schwope}, {Wilms}, \& {German eROSITA Consortium}}]{erosita}
{Merloni}, A., {Predehl}, P., {Becker}, W., {et~al.} 2012, arXiv e-prints, arXiv:1209.3114, \dodoi{10.48550/arXiv.1209.3114}

\bibitem[{{Motte} {et~al.}(2003){Motte}, {Schilke}, \& {Lis}}]{motte03}
{Motte}, F., {Schilke}, P., \& {Lis}, D.~C. 2003, \apj, 582, 277, \dodoi{10.1086/344538}

\bibitem[{{Nguyen Luong} {et~al.}(2011){Nguyen Luong}, {Motte}, {Schuller}, {Schneider}, {Bontemps}, {Schilke}, {Menten}, {Heitsch}, {Wyrowski}, {Carlhoff}, {Bronfman}, \& {Henning}}]{luong11}
{Nguyen Luong}, Q., {Motte}, F., {Schuller}, F., {et~al.} 2011, \aap, 529, A41, \dodoi{10.1051/0004-6361/201016271}

\bibitem[{{Planck Collaboration} {et~al.}(2014){Planck Collaboration}, {Abergel}, {Ade}, {Aghanim}, {Alves}, {Aniano}, {Armitage-Caplan}, {Arnaud}, {Ashdown}, {Atrio-Barandela}, {Aumont}, {Baccigalupi}, {Banday}, {Barreiro}, {Bartlett}, {Battaner}, {Benabed}, {Beno{\^\i}t}, {Benoit-L{\'e}vy}, {Bernard}, {Bersanelli}, {Bielewicz}, {Bobin}, {Bock}, {Bonaldi}, {Bond}, {Borrill}, {Bouchet}, {Boulanger}, {Bridges}, {Bucher}, {Burigana}, {Butler}, {Cardoso}, {Catalano}, {Chamballu}, {Chary}, {Chiang}, {Chiang}, {Christensen}, {Church}, {Clemens}, {Clements}, {Colombi}, {Colombo}, {Combet}, {Couchot}, {Coulais}, {Crill}, {Curto}, {Cuttaia}, {Danese}, {Davies}, {Davis}, {de Bernardis}, {de Rosa}, {de Zotti}, {Delabrouille}, {Delouis}, {D{\'e}sert}, {Dickinson}, {Diego}, {Dole}, {Donzelli}, {Dor{\'e}}, {Douspis}, {Draine}, {Dupac}, {Efstathiou}, {En{\ss}lin}, {Eriksen}, {Falgarone}, {Finelli}, {Forni}, {Frailis}, {Fraisse}, {Franceschi}, {Galeotta}, {Ganga}, {Ghosh}, {Giard}, {Giardino}, {Giraud-H{\'e}raud},
  {Gonz{\'a}lez-Nuevo}, {G{\'o}rski}, {Gratton}, {Gregorio}, {Grenier}, {Gruppuso}, {Guillet}, {Hansen}, {Hanson}, {Harrison}, {Helou}, {Henrot-Versill{\'e}}, {Hern{\'a}ndez-Monteagudo}, {Herranz}, {Hildebrandt}, {Hivon}, {Hobson}, {Holmes}, {Hornstrup}, {Hovest}, {Huffenberger}, {Jaffe}, {Jaffe}, {Jewell}, {Joncas}, {Jones}, {Juvela}, {Keih{\"a}nen}, {Keskitalo}, {Kisner}, {Knoche}, {Knox}, {Kunz}, {Kurki-Suonio}, {Lagache}, {L{\"a}hteenm{\"a}ki}, {Lamarre}, {Lasenby}, {Laureijs}, {Lawrence}, {Leonardi}, {Le{\'o}n-Tavares}, {Lesgourgues}, {Levrier}, {Liguori}, {Lilje}, {Linden-V{\o}rnle}, {L{\'o}pez-Caniego}, {Lubin}, {Mac{\'\i}as-P{\'e}rez}, {Maffei}, {Maino}, {Mandolesi}, {Maris}, {Marshall}, {Martin}, {Mart{\'\i}nez-Gonz{\'a}lez}, {Masi}, {Massardi}, {Matarrese}, {Matthai}, {Mazzotta}, {McGehee}, {Melchiorri}, {Mendes}, {Mennella}, {Migliaccio}, {Mitra}, {Miville-Desch{\^e}nes}, {Moneti}, {Montier}, {Morgante}, {Mortlock}, {Munshi}, {Murphy}, {Naselsky}, {Nati}, {Natoli}, {Netterfield},
  {N{\o}rgaard-Nielsen}, {Noviello}, {Novikov}, {Novikov}, {Osborne}, {Oxborrow}, {Paci}, {Pagano}, {Pajot}, {Paladini}, {Paoletti}, {Pasian}, {Patanchon}, {Perdereau}, {Perotto}, {Perrotta}, {Piacentini}, {Piat}, {Pierpaoli}, {Pietrobon}, {Plaszczynski}, {Pointecouteau}, {Polenta}, {Ponthieu}, {Popa}, {Poutanen}, {Pratt}, {Pr{\'e}zeau}, {Prunet}, {Puget}, {Rachen}, {Reach}, {Rebolo}, {Reinecke}, {Remazeilles}, {Renault}, {Ricciardi}, {Riller}, {Ristorcelli}, {Rocha}, {Rosset}, {Roudier}, {Rowan-Robinson}, {Rubi{\~n}o-Mart{\'\i}n}, {Rusholme}, {Sandri}, {Santos}, {Savini}, {Scott}, {Seiffert}, {Shellard}, {Spencer}, {Starck}, {Stolyarov}, {Stompor}, {Sudiwala}, {Sunyaev}, {Sureau}, {Sutton}, {Suur-Uski}, {Sygnet}, {Tauber}, {Tavagnacco}, {Terenzi}, {Toffolatti}, {Tomasi}, {Tristram}, {Tucci}, {Tuovinen}, {T{\"u}rler}, {Umana}, {Valenziano}, {Valiviita}, {Van Tent}, {Verstraete}, {Vielva}, {Villa}, {Vittorio}, {Wade}, {Wandelt}, {Welikala}, {Ysard}, {Yvon}, {Zacchei}, \& {Zonca}}]{planck_dust2}
{Planck Collaboration}, {Abergel}, A., {Ade}, P.~A.~R., {et~al.} 2014, \aap, 571, A11, \dodoi{10.1051/0004-6361/201323195}

\bibitem[{{Planck Collaboration} {et~al.}(2016){Planck Collaboration}, {Aghanim}, {Ashdown}, {Aumont}, {Baccigalupi}, {Ballardini}, {Banday}, {Barreiro}, {Bartolo}, {Basak}, {Benabed}, {Bernard}, {Bersanelli}, {Bielewicz}, {Bonavera}, {Bond}, {Borrill}, {Bouchet}, {Boulanger}, {Burigana}, {Calabrese}, {Cardoso}, {Carron}, {Chiang}, {Colombo}, {Comis}, {Couchot}, {Coulais}, {Crill}, {Curto}, {Cuttaia}, {de Bernardis}, {de Zotti}, {Delabrouille}, {Di Valentino}, {Dickinson}, {Diego}, {Dor{\'e}}, {Douspis}, {Ducout}, {Dupac}, {Dusini}, {Elsner}, {En{\ss}lin}, {Eriksen}, {Falgarone}, {Fantaye}, {Finelli}, {Forastieri}, {Frailis}, {Fraisse}, {Franceschi}, {Frolov}, {Galeotta}, {Galli}, {Ganga}, {G{\'e}nova-Santos}, {Gerbino}, {Ghosh}, {Giraud-H{\'e}raud}, {Gonz{\'a}lez-Nuevo}, {G{\'o}rski}, {Gruppuso}, {Gudmundsson}, {Hansen}, {Helou}, {Henrot-Versill{\'e}}, {Herranz}, {Hivon}, {Huang}, {Jaffe}, {Jones}, {Keih{\"a}nen}, {Keskitalo}, {Kiiveri}, {Kisner}, {Krachmalnicoff}, {Kunz}, {Kurki-Suonio}, {Lamarre},
  {Langer}, {Lasenby}, {Lattanzi}, {Lawrence}, {Le Jeune}, {Levrier}, {Lilje}, {Lilley}, {Lindholm}, {L{\'o}pez-Caniego}, {Ma}, {Mac{\'\i}as-P{\'e}rez}, {Maggio}, {Maino}, {Mandolesi}, {Mangilli}, {Maris}, {Martin}, {Mart{\'\i}nez-Gonz{\'a}lez}, {Matarrese}, {Mauri}, {McEwen}, {Melchiorri}, {Mennella}, {Migliaccio}, {Miville-Desch{\^e}nes}, {Molinari}, {Moneti}, {Montier}, {Morgante}, {Moss}, {Natoli}, {Oxborrow}, {Pagano}, {Paoletti}, {Patanchon}, {Perdereau}, {Perotto}, {Pettorino}, {Piacentini}, {Plaszczynski}, {Polastri}, {Polenta}, {Puget}, {Rachen}, {Racine}, {Reinecke}, {Remazeilles}, {Renzi}, {Rocha}, {Rosset}, {Rossetti}, {Roudier}, {Rubi{\~n}o-Mart{\'\i}n}, {Ruiz-Granados}, {Salvati}, {Sandri}, {Savelainen}, {Scott}, {Sirignano}, {Sirri}, {Soler}, {Spencer}, {Suur-Uski}, {Tauber}, {Tavagnacco}, {Tenti}, {Toffolatti}, {Tomasi}, {Tristram}, {Trombetti}, {Valiviita}, {Van Tent}, {Vielva}, {Villa}, {Vittorio}, {Wandelt}, {Wehus}, {Zacchei}, \& {Zonca}}]{planck_dust1}
{Planck Collaboration}, {Aghanim}, N., {Ashdown}, M., {et~al.} 2016, \aap, 596, A109, \dodoi{10.1051/0004-6361/201629022}

\bibitem[{Popescu {et~al.}(2017)Popescu, Yang, Tuffs, Natale, Rushton, \& Aharonian}]{popescu17}
Popescu, C.~C., Yang, R., Tuffs, R.~J., {et~al.} 2017, Monthly Notices of the Royal Astronomical Society, 470, 2539, \dodoi{10.1093/mnras/stx1282}

\bibitem[{{Smith} {et~al.}(1978){Smith}, {Biermann}, \& {Mezger}}]{smith78}
{Smith}, L.~F., {Biermann}, P., \& {Mezger}, P.~G. 1978, \aap, 66, 65

\bibitem[{{Umemoto} {et~al.}(2017){Umemoto}, {Minamidani}, {Kuno}, {Fujita}, {Matsuo}, {Nishimura}, {Torii}, {Tosaki}, {Kohno}, {Kuriki}, {Tsuda}, {Hirota}, {Ohashi}, {Yamagishi}, {Handa}, {Nakanishi}, {Omodaka}, {Koide}, {Matsumoto}, {Onishi}, {Tokuda}, {Seta}, {Kobayashi}, {Tachihara}, {Sano}, {Hattori}, {Onodera}, {Oasa}, {Kamegai}, {Tsuboi}, {Sofue}, {Higuchi}, {Chibueze}, {Mizuno}, {Honma}, {Muller}, {Inoue}, {Morokuma-Matsui}, {Shinnaga}, {Ozawa}, {Takahashi}, {Yoshiike}, {Costes}, \& {Kuwahara}}]{Umemoto2017}
{Umemoto}, T., {Minamidani}, T., {Kuno}, N., {et~al.} 2017, \pasj, 69, 78, \dodoi{10.1093/pasj/psx061}

\bibitem[{{Vercellone} {et~al.}(2022){Vercellone}, {Bigongiari}, {Burtovoi}, {Cardillo}, {Catalano}, {Franceschini}, {Lombardi}, {Nava}, {Pintore}, {Stamerra}, {Tavecchio}, {Zampieri}, {Alves Batista}, {Amato}, {Antonelli}, {Arcaro}, {Becerra Gonz{\'a}lez}, {Bonnoli}, {B{\"o}ttcher}, {Brunetti}, {Compagnino}, {Crestan}, {D'A{\`\i}}, {Fiori}, {Galanti}, {Giuliani}, {de Gouveia Dal Pino}, {Green}, {Lamastra}, {Landoni}, {Lucarelli}, {Morlino}, {Olmi}, {Peretti}, {Piano}, {Ponti}, {Poretti}, {Romano}, {Saturni}, {Scuderi}, {Tutone}, {Umana}, {Acosta-Pulido}, {Barai}, {Bonanno}, {Bonanno}, {Bruno}, {Bulgarelli}, {Conforti}, {Costa}, {Cusumano}, {Del Santo}, {del Valle}, {Della Ceca}, {Falceta-Gon{\c{c}}alves}, {Fioretti}, {Germani}, {Garc{\'\i}a-L{\'o}pez}, {Ghedina}, {Gianotti}, {Giordano}, {Kreter}, {Incardona}, {Iovenitti}, {La Barbera}, {La Palombara}, {La Parola}, {Leto}, {Longo}, {L{\'o}pez-Oramas}, {Maccarone}, {Mereghetti}, {Millul}, {Naletto}, {Pagliaro}, {Parmiggiani}, {Righi},
  {Rodr{\'\i}guez-Ram{\'\i}rez}, {Romeo}, {Sangiorgi}, {Santos de Lima}, {Tagliaferri}, {Testa}, {Tosti}, {V{\'a}zquez Acosta}, {{\.Z}ywucka}, {Caraveo}, \& {Pareschi}}]{astri}
{Vercellone}, S., {Bigongiari}, C., {Burtovoi}, A., {et~al.} 2022, Journal of High Energy Astrophysics, 35, 1, \dodoi{10.1016/j.jheap.2022.05.005}

\bibitem[{{Vieu} {et~al.}(2022){Vieu}, {Gabici}, {Tatischeff}, \& {Ravikularaman}}]{vieu22}
{Vieu}, T., {Gabici}, S., {Tatischeff}, V., \& {Ravikularaman}, S. 2022, \mnras, 512, 1275, \dodoi{10.1093/mnras/stac543}

\bibitem[{{Wang} {et~al.}(2020){Wang}, {Beuther}, {Rugel}, {Soler}, {Stil}, {Ott}, {Bihr}, {McClure-Griffiths}, {Anderson}, {Klessen}, {Goldsmith}, {Roy}, {Glover}, {Urquhart}, {Heyer}, {Linz}, {Smith}, {Bigiel}, {Dempsey}, \& {Henning}}]{Wang2019}
{Wang}, Y., {Beuther}, H., {Rugel}, M.~R., {et~al.} 2020, \aap, 634, A83, \dodoi{10.1051/0004-6361/201937095}

\bibitem[{Wilks(1938)}]{wilks}
Wilks, S.~S. 1938, The Annals of Mathematical Statistics, 9, 60 , \dodoi{10.1214/aoms/1177732360}

\bibitem[{{Yang} \& {Wang}(2020)}]{yang20}
{Yang}, R.-Z., \& {Wang}, Y. 2020, \aap, 640, A60, \dodoi{10.1051/0004-6361/202037518}

\bibitem[{{Yuan} {et~al.}(2022){Yuan}, {Zhang}, {Chen}, \& {Ling}}]{ep}
{Yuan}, W., {Zhang}, C., {Chen}, Y., \& {Ling}, Z. 2022, in Handbook of X-ray and Gamma-ray Astrophysics, 86, \dodoi{10.1007/978-981-16-4544-0_151-1}

\bibitem[{{Zabalza}(2015)}]{naima}
{Zabalza}, V. 2015, in International Cosmic Ray Conference, Vol.~34, 34th International Cosmic Ray Conference (ICRC2015), 922.
\newblock \doarXiv{1509.03319}

\bibitem[{{Zhang} {et~al.}(2014){Zhang}, {Moscadelli}, {Sato}, {Reid}, {Menten}, {Zheng}, {Brunthaler}, {Dame}, {Xu}, \& {Immer}}]{w43_distance}
{Zhang}, B., {Moscadelli}, L., {Sato}, M., {et~al.} 2014, \apj, 781, 89, \dodoi{10.1088/0004-637X/781/2/89}

\bibitem[{Zhang(2023)}]{lact}
Zhang, S. 2023, in Proceedings of 38th International Cosmic Ray Conference {\textemdash} PoS(ICRC2023), Vol. 444, 808, \dodoi{10.22323/1.444.0808}

\end{thebibliography}
